\documentclass[aps,pra,twocolumn,amsmath,amssymb,nofootinbib,showpacs,superscriptaddress]{revtex4-1}
\usepackage[english]{babel}
    \usepackage[T1]{fontenc}
    \usepackage{ae}
    \usepackage{aecompl}
\usepackage{latexsym}
\usepackage{graphics}
\usepackage[epsfig]{graphicx}
\usepackage{epsfig}
\usepackage{color}
\usepackage{bm}
\usepackage{amsmath}
\usepackage{amssymb}
\usepackage{amsthm}
\usepackage{dcolumn}
\usepackage{bm}
\usepackage{float}
\usepackage[hyperindex,breaklinks]{hyperref}
\usepackage{color}
\usepackage{epstopdf}
\usepackage{cleveref}
\usepackage[svgnames]{xcolor}
\usepackage{enumerate}
\usepackage{braket}
\hypersetup{hidelinks,colorlinks=true,allcolors=DarkBlue}

\DeclareMathOperator{\Tr}{Tr}

\theoremstyle{remark}
\newtheorem{remark}{Remark}
\theoremstyle{plain}
\newtheorem{theorem}{Theorem}
\newtheorem{corollary}{Corollary}

\begin{document}

\preprint{APS/123-QED}

\title{Security of quantum key distribution with detection-efficiency mismatch in the single-photon case: Tight bounds}

\author{M. K. Bochkov} 
\affiliation{National Research Nuclear University MEPhI, Moscow 115409, Russia}

\author{A. S. Trushechkin}
\affiliation{National Research Nuclear University MEPhI, Moscow 115409, Russia}
\affiliation{Steklov Mathematical Institute of Russian Academy of Sciences, Moscow 119991, Russia}
\affiliation{Russian Quantum Center, Skolkovo, Moscow 143025, Russia}
\affiliation{National University of Science and Technology MISiS, Moscow 119049, Russia}

\date{\today}
\begin{abstract}
One of the challenges in practical quantum key distribution is dealing with efficiency mismatch between different threshold single-photon detectors. There are known bounds for the secret key rate for the BB84 protocol with detection-efficiency mismatch provided that the eavesdropper sends no more than one photon to the legitimate receiver. Here we improve these bounds and give tight bounds for the secret key rate with a constant detection-efficiency mismatch under the same single-photon assumption. We propose a method based on the analytical minimization of the relative entropy of coherence, which can be used in other problems in quantum key distribution. Also we propose an adaptation of the decoy state method to proof the security in the case of weak coherent pulses  on the source side.
\end{abstract}
\maketitle

\section{Introduction}

Quantum key distribution (QKD) is a way for two distant parties (Alice and Bob, the legitimate parties) to establish a common secret key for confidential messaging. Theoretically, the security of QKD is based solely on the laws of quantum mechanics, i.e., does not depend on the computational power or technical devices of an eavesdropper (Eve). However, in practice QKD faces  certain challenges caused by imperfect devices \cite{Gisin,Scarani,nature2016}. One of such imperfections is  efficiency mismatch between different threshold single-photon detectors. 

In the most common QKD protocol,  Bennett and Brassard 1984 (BB84), as well as in other discrete-variable QKD protocols, information is typically encoded in the polarization or phase of weak coherent pulses simulating true single-photon states. Hence, the corresponding implementations employ single-photon detection techniques. Ideally, a detector should click whenever it is hit by at least one photon. However, a realistic detector is triggered by a photon only with a certain probability  $0<\eta<1$, which is referred to as the efficiency of a detector. Typical value of $\eta$ for the detectors used in practical QKD systems (based on avalanche photodiodes) is 0.1. The detectors based on superconductors have $\eta\approx0.9$, but they are more expensive and require cryogenic temperatures.

In this paper we will consider the BB84 protocol with the active basis choice. In this case, Bob uses two detectors: One  for the signals encoding bit 0 and one for the signals encoding bit 1, respectively. If both detectors have the same efficiency $\eta$, then the loss in the detection rate can be treated as a part of transmission loss followed by ideal detectors with perfect efficiency. However, in practice, it is  hard to build two detectors with exactly the same efficiencies. So, the problem of detector-efficiency mismatch arises. In this case we cannot  treat the detection loss as a part of the transmission loss anymore since  the detection loss is different for different detectors. Also, in general, usual proofs of security of QKD \cite{ShorPreskill, Renner,GisinRenner,Toma} are not applicable to this case.

For example, if detection-efficiency mismatch takes place, then the frequency of, for example, zeros is greater then the frequency of ones in the raw key. This increases  Eve's \textit{a priori} information on the raw key and, hence, require larger key contraction on the privacy amplification step. 

The situation becomes even more complicated if Eve has ability to control the efficiencies in some way (for example, by manipulation with spatial modes) \cite{Fung}. Such attacks are described and employed experimentally \cite{LoHack,Makarov}. Under some conditions,  Eve can completely control  Bob's measurements and obtain full information on the secret key.

The security of the BB84 protocol with  detection-efficiency mismatch is proved in Ref.~\cite{Fung}, but the proof is restricted to the case when exactly one photon may arrive at  Bob's side. Here we improve the bounds and give tight bounds for the secret key rate with a constant detection-efficiency mismatch provided that Eve cannot send more than one photon to Bob. Here we do not address the case when Eve can control the efficiencies of the detectors and consider only the case of a constant detection-efficiency mismatch.

We adopt the approach introduced in Refs.~\cite{Lutk-numeric,Lutk-num-pre} based on the minimization of the relative entropy of coherence for the purpose of numerical determination of the secret key rate. We show that the analytical (rather than numerical) minimization of the relative entropy of coherence also can be used as a method of solving  QKD problems.

The text is organized as follows. In Sec.~\ref{SecPM}, we formulate a prepare-and-measure version of the BB84 protocol and specify Bob's positive operator-valued measure (POVM) for the case of detection-efficiency mismatch. In Sec.~\ref{SecEnt}, we give an equivalent entanglement-based formulation of the protocol. In Sec.~\ref{SecKeyRate}, we review the approach of Refs.~\cite{Lutk-numeric,Lutk-num-pre}, which reduces the calculation of the secret key rate to a convex optimization problem, and state a theorem with the analytic formula for the secret key rate, Eq.~(\ref{EqFin}). A slight modification of this formula (\ref{EqFin2}), which outperforms (\ref{EqFin}) if the detection-efficiency mismatch or quantum bit error rate (QBER) is large, is given after the theorem. The proof of the theorem is given in Appendix~A, and the leakage of information in the error correction procedure for the case of detection-efficiency mismatch is analyzed in Appendix~B. Finally, in Sec.~\ref{SecDecoy},  we propose an adaptation of the decoy state method to prove the security of the BB84 protocol with detection-efficiency mismatch in the case of weak coherent pulses  on the source side.

\section{Prepare-and-measure formulation of the BB84 protocol with detection-efficiency mismatch}\label{SecPM}

We start with the description of a mathematical model of the BB84 protocol with detection-efficiency mismatch. Most practical implementations of the BB84 protocol are  prepare-and-measure based, in which Alice sends quantum states to Bob. We assume that Alice sends true qubits to Bob, i.e., single-photon pulses with  information encoded in some two-dimensional variable. Hence, Alice's Hilbert space is $\mathcal H_A=\mathbb C^2$. We will use two bases of $\mathbb C^2$: the standard one ($z$ basis) $\{\ket0,\ket1\}$ and the Hadamard one ($x$ basis) $\{\ket+,\ket-\}$, $\ket\pm=(\ket0\pm\ket1)/\sqrt2$. In each basis, the first element encodes the bit 0, and the second element encodes the bit 1.

\begin{remark}
In most implementations, Alice sends not true single-photon states but weak coherent pulses, which make QKD vulnerable to the photon number splitting attack \cite{PNS,PNS2}. However, this problem can be fixed by the decoy-state method, which  effectively allows us to bound the number of multiphoton pulses from above \cite{LoMa2005,Wang2005,MaLo2005,Ma2017,Trushechkin2016}. These pulses  are treated as insecure, i.e., the information encoded in such pulses is assumed to be known to Eve. After the estimation of the number of multiphoton pulses, it suffices to bound  Eve's information on the key bits originated from  the single-photon pulses.

The decoy state method for the detector-efficiency mismatch case is considered in Ref.~\cite{DecoyMismatch} with an additional symmetry assumption that the so-called yields of $i$-photon states ($i=0,1,2,\ldots$) do not depend on the measurement basis. This is true for the case of identical detectors, but generally not true in the case of detection-efficiency mismatch. We propose a realization of the decoy state method for the case of detection-efficiency mismatch without such assumption later in Sec.~\ref{SecDecoy}. As we will see, we need more detailed data than in the case of identical detectors. In this and in the next two sections we assume that Alice sends true single-photon pulses.
\end{remark}

Bob measures the signals in an infinite-dimensional mode space with no limit on the number of photons. Eve can use this fact for her advantage: She can send to Bob as many photons as she wishes.  The issues with the number of photons can be avoided in QKD with perfect detectors due to the squashing model method \cite{squash}. However, there is no squashing model for the detection-efficiency mismatch case, hence, zero-photon and multiphoton cases should be analyzed explicitly.  As in  Ref.~\cite{Fung}, our security proof is based on the additional assumption that Eve sends no more than one photon to Bob. Note that a possible way to assure that Eve does not send more than one photon to Bob is the use of the detector-decoy idea \cite{DetDecoy}.  Another approach to bound the multiphoton pulses is proposed in Ref.~\cite{LutkEnt}.

Thus, the Bob's Hilbert space is $\mathcal H_B=\mathbb C^3$ and is spanned by three vectors: $\ket0,\ket1$, and $\ket{\rm vac}$ (a vacuum vector).

Let us describe the Bob's POVM. Bob chooses the $z$ measurement basis with the probability $p_z$ and the $x$ basis with the probability $p_x=1-p_z$. We consider the BB84 protocol with single-photon detectors and the active basis choice. In this case, Bob uses two single-photon detectors: One for the signals encoding bit 0 and one for the signals encoding bit 1. 

Let the efficiencies of the Bob's detectors be $\eta_0$ and $\eta_1\neq\eta_0$, respectively. Let, for definiteness, $1\geq\eta_0>\eta_1>0$. Then the efficiencies of the detectors can be renormalized as $\eta'_0=1$ and $\eta'_1=\eta=\eta_1/\eta_0$, and the common loss $\eta_0$  in both detectors can be treated as  additional transmission loss \cite{LutkEnt}. The mismatch parameter $\eta$ is assumed to be constant and known to both legitimate parties and the eavesdropper.

As shown in Ref.~\cite{squash}, without loss of generality, we can think that the actual measurement is preceded by a quantum non-demolition (QND) measurement of the  number of photons (POVM $\{\ket{\rm vac}\bra{\rm vac}, I_2\}$, where $I_2$ is a identity operator in the two-dimensional single-photon subspace, which is spanned by $\ket0$ and $\ket1$). Then, the Bob's POVM is as follows:
\begin{subequations}\label{EqPOVMb}
\begin{eqnarray}
P^B_{z,0}&=p_z\ket0\bra0,\quad &P^B_{z,1}=p_z\eta\ket1\bra1,\\
P^B_{x,0}&=p_x\ket+\bra+,\quad &P^B_{x,1}=p_x\eta\ket-\bra-,\\
P^B_\varnothing&=I_3-P^B_{z,0}-P^B_{z,1}&-P^B_{x,0}-P^B_{x,1},
\end{eqnarray}
\end{subequations}
where $I_3$ is the identity operator in the three-dimensional Bob's space and $\varnothing$ corresponds to the outcome ``no click.'' It happens whenever either the outcome of the QND is ${\rm vac}$ or a photon hits the detector 1, but the no-click event is activated with the probability $1-\eta$.

Now we describe the protocol. 

\begin{enumerate}
\item Alice randomly, with the probabilities $(1/2,1/2)$, chooses a bit value $\overline a\in\{0,1\}$;

\item Alice randomly, with the probabilities $(p_z,p_x=1-p_z)$, chooses a basis: either the $z$ basis or $x$ basis. It is assumed that $p_z\approx1$. Only the $z$ basis is used for the key generation, while the $x$ basis is used only for the detection of eavesdropping (see below).

\item Bob also chooses a measurement basis: either the $z$ basis or $x$ basis, also with the probabilities $(p_z,p_x)$.

\item Alice generates a photon in the state depending on the basis and the bit value and send it to Bob. For example, if Alice has chosen the bit value 0 and the $x$ basis, she sends a photon in the state $\ket+$. Bob measures this photon according to  POVM (\ref{EqPOVMb}) and, if at least one detector clicks, obtain the bit value $\overline b$.

\item Alice and Bob repeat steps 1--3 a large number of times, $n$. As a result, they have their own bit strings $\overline{\mathbf a}$ and $\overline{\mathbf b}$, which are referred to as the \textit{raw keys}. 

\item \textit{Announcements:} Bob announces the numbers of positions where he has obtained a click over a public authentic classical channel. Alice and Bob announce the bases they used and the bit values for positions where both used the $x$ basis.

\item \textit{Sifting:} Alice and Bob keep the states where they both chose the $z$ basis and Bob obtained a click. The other positions are dropped from the raw keys. The resulting keys are referred to as the \textit{sifted keys}.

\item \textit{Parameter estimation:} Alice and Bob analyze the announced data and estimate the amount of information on the Alice's sifted key that can be known to Eve. If this amount is not too large, they continue. Otherwise, they abort the protocol.

\item \textit{Error correction:} The mismatches between the bit values of Alice's and Bob's sifted keys are treated as the errors in the Bob's key.  Alice sends to Bob a syndrome of her sifted key over the classical authentic channel. Using the syndrome, Bob corrects the errors in his key. Now the Bob's corrected key coincides with the Alice's sifted key.

\item \textit{Privacy amplification:} Alice sends to Bob a hash function, and both  apply this function to their (identical) keys. The hash function maps the key to a shorter key about which Eve has only negligible amount of information. This is the \textit{final key}, or \textit{secret key}.
\end{enumerate}

\begin{remark}
In this paper we consider only asymptotic case $n\to\infty$ and do not address the finite-key effects \cite{Renner,GisinRenner}. For this reason, we also do not address some practical peculiarities like the verification after the error-correction step  \cite{Zbinden2014,Fedorov2016} or the use of both bases for the key generation since they do not matter in the asymptotic case.
\end{remark}

\section{Entanglement-based formulation of the protocol}\label{SecEnt}

We have described the prepare-and-measure implementation of the BB84 protocol, which is the most common in practice. However, a common mathematical trick is to reformulate the protocol in an equivalent  entanglement-based version. In the entanglement-based version of the protocol, an entangled state $\rho_{AB}$ is distributed among the legitimate parties.  Alice's measurement result encodes the information about which state she prepared. Let us describe the entanglement-based version of the BB84 protocol.

The steps 1, 2, and 4 of the protocol given above are replaced by the following:

\begin{enumerate}
\item A source of entangled states generates a two-qubit entangled state 
\begin{equation}\label{EqPhi}
\rho_{AB}=\ket\Phi_{AB}\bra\Phi,
\end{equation}
where 
\begin{equation}
\ket\Phi_{AB} =\frac1{\sqrt2}(\ket{0}_A\otimes\ket{0}_B+\ket{1}_A\otimes\ket{1}_B),
\end{equation}
and sends the first qubit to Alice and the second qubit to Bob.

\item Alice perform a measurement of her part according to the POVM
\begin{equation}
\begin{split}
P^A_{z,0}&=p_z\ket0\bra0,\quad P^A_{z,1}=p_z\ket1\bra1,\\
P^A_{x,0}&=p_x\ket+\bra+,\quad P^A_{x,1}=p_x\ket-\bra-.
\end{split}
\end{equation}

\end{enumerate}

The probabilities of obtaining these results provided that the state is $\rho_A=\Tr_B\rho_{AB}=I_2/2$ are: $p_z/2,p_z/2,p_x/2$, and $p_x/2$, respectively. If Alice obtains the result, say $(x,1)$, then the state is changed to
\begin{equation}
\rho_{AB}\to \ket{-}_A\bra-\otimes\ket-_B\bra{-},
\end{equation}
i.e., is equivalent to sending the state $\ket-$ to Bob. In this sense, the entanglement-based formulation is mathematically equivalent to the prepare-and-measure one. Note that since Alice's measurement is virtual, it corresponds to detectors with  perfect efficiency. 

In the prepare-and-measure formulation of the protocol, Eve controls the transmission channel between Alice and Bob. In the equivalent entanglement-based  formulation this is modeled by the Eve's ability to replace (\ref{EqPhi}) by her own density operator $\rho_{AB}$ acting on the space $\mathcal H_A\otimes\mathcal H_B=\mathbb C^2\otimes\mathbb C^3$ under the restriction of the fixed $\rho_A=\Tr_B\rho_{AB}$.  This means that the subsystem $A$ is inaccessible for Eve. One may think about the source of entangled states placed inside the Alice's laboratory, so, Eve has access only to the subsystem $B$ transmitting over the channel. In our case (\ref{EqPhi}) we have  $\rho_{A}=I_2/2$.

\section{Secret key rate}\label{SecKeyRate}

We  define the secret key rate as the ratio of the length of the final key to the number of  channel uses $n$. To derive the formula for it, we need to formalize the steps of the QKD protocol given above. We adopt the mathematical model developed in Refs.~\cite{Lutk-numeric,Lutk-num-pre}.

Alice's and Bob's measurements with the announcements are described by the following quantum channel:

\begin{equation}
\begin{split}
\rho_{AB}&\mapsto \sum_{a\in\{z,x\}}\sum_{b\in\{z,x\}}
K_a^A\otimes K_b^B\rho_{AB}(K_a^A\otimes K_b^B)^\dag\\&+
\sum_{a\in\{z,x\}}K_a^A\otimes K_\varnothing^B\rho_{AB}(K_a^A\otimes K_\varnothing^B)^\dag\\
&=\mathcal M(\rho_{AB})=
\rho^{(2)}_{A\widetilde A\overline AB\widetilde B\overline B},
\end{split}
\end{equation}
where
\begin{subequations}
\begin{eqnarray}
K_a^A&=&\sum_{\alpha\in\{0,1\}}
\sqrt{P^A_{a,\alpha}}\otimes\ket a_{\widetilde A}
\otimes\ket \alpha_{\overline A},\\
K_b^B&=&\sum_{\beta\in\{0,1\}}
\sqrt{P^B_{b,\beta}}\otimes\ket b_{\widetilde B}
\otimes\ket \beta_{\overline B},\\
K_\varnothing^B&=&\sqrt{P^B_\varnothing}
\ket{\varnothing}_{\widetilde B}\ket 0_{\overline B}.
\end{eqnarray}
\end{subequations}
Here the registers $\widetilde A$ (two-dimensional) and $\widetilde B$ (three-dimensional) store the information that is announced in the public channel: bases choices and the result $\varnothing$ for Bob. The two-dimensional registers $\overline A$ and $\overline B$ store the key bit, which is not announced.

The sifting step is described by the projector
\begin{equation}
\Pi=\ket z_{\widetilde A}\bra z\otimes \ket z_{\widetilde B}\bra z,
\end{equation}
\begin{equation}
\rho^{(2)}_{A\widetilde A\overline AB\widetilde B\overline B}\mapsto
\frac1{p_{\rm pass}}
\Pi\rho^{(2)}_{A\widetilde A\overline AB\widetilde B\overline B}\Pi=
\rho^{(3)}_{A\widetilde A\overline AB\widetilde B\overline B},
\end{equation}
where $p_{\rm pass}=\Tr\Pi\rho^{(2)}_{A\widetilde A\overline AB\widetilde B\overline B}$ is the probability of passing the sifting stage.

In the general scheme given in Ref.~\cite{Lutk-numeric}, an additional key map $g(a,b,\alpha)$ is used to form Alice's key bit in an additional register. In this particular protocol, we do not need this since  Alice's and Bob's key bits are already given in the registers $\overline A$ and $\overline B$. Let us also define a pinching (partially dephasing) quantum channel 
\begin{equation}
\mathcal Z(\sigma)
=\sum_{\alpha\in\{0,1\}}(\ket \alpha_{\overline A}\bra \alpha\otimes I)\,\sigma\,
(\ket \alpha_{\overline A}\bra \alpha\otimes I),
\end{equation}
where $I$ is the identity operator acting on all remaining spaces except $\overline A$, and 
\begin{equation}
\mathcal Z
(\rho^{(3)}_{A\widetilde A\overline AB\widetilde B\overline B})=
\rho^{(4)}_{A\widetilde A\overline AB\widetilde B\overline B}.
\end{equation}

As we said in the end of Sec.~\ref{SecEnt}, Eve can choose the state $\rho_{AB}$ under the restriction $\rho_A=\Tr_B\rho_{AB}=I_2/2$. Also, according to the purification theorem, $\rho_{AB}$ can be expressed as $\rho_{AB}=\Tr_E\rho_{ABE}$, where $\rho_{ABE}$ is a pure state in a larger Hilbert space $\mathcal H_A\otimes\mathcal H_B\otimes\mathcal H_E$, and all such representations are unitary equivalent \cite{Holevo}. Eve is assumed to own the additional register $E$ (the purification of $\rho_{AB}$).

In this paper we restrict our analysis to the case when Eve performs a collective attack: She prepares $n$ equal copies of $\rho_{ABE}$ and then perform a collective measurement on her parts $E$ in all copies. According to Devetak and Winter theorem \cite{DW}, the asymptotic ($n\to\infty$) secret key rate is given by

\begin{equation}\label{EqDW}
K=p_{\rm pass}\left[
H(\overline A|E\widetilde A\widetilde B)_{\rho^{(4)}}-
H(\overline A|\overline B\widetilde A\widetilde B)_{\rho^{(4)}}
\right],
\end{equation}
where $H$ is the conditional von Neumann entropy.
Here the first and the second terms in the brackets characterize  Eve's and  Bob's ignorances about  Alice's key bit, respectively. Here we assume that the length of the error-correcting syndrome is given by the Shannon theoretical limit. Otherwise, a factor $f>1$ should be added to the second term. The present-day error-correcting codes allow for $f=1.22$. A  method of  using of the low-density parity-check codes in QKD, which allows us to decrease the factor $f$, is given in Refs.~\cite{SymBl1,SymBl2}. A  syndrome-based QBER estimation algorithm, which also can decrease $f$, is proposed in Ref.~\cite{fly}.

The second term in the right-hand side of Eq.~(\ref{EqDW}) is bounded from above by $h(Q)$, where $h$ is the binary entropy and $Q$ is the QBER in the $z$ basis. Note that, due to detection-efficiency mismatch, the channel from $\overline A$ to $\overline B$ is not a binary symmetric channel. However, in our case the second term can be taken to be equal to $h(Q)$; see Appendix~B for details. The QBER is a value observed by Alice and Bob. 

Eve's ignorance should be estimated from below. By Theorem~1 from Ref.~\cite{Coles},
\begin{eqnarray}
H(\overline A|E\widetilde A\widetilde B)_{\rho^{(4)}}&=&
D(\rho^{(3)}_{A\widetilde A\overline AB\widetilde B\overline B}\|
\rho^{(4)}_{A\widetilde A\overline AB\widetilde B\overline B})\nonumber\\
&=&p_{\rm pass}^{-1}D\big(\mathcal G(\rho_{AB})\|\mathcal Z(\mathcal G(\rho_{AB}))\big),\quad\label{EqD}
\end{eqnarray}
where 
\begin{equation}
\mathcal G(\rho_{AB})=\Pi\mathcal M(\rho_{AB})\Pi
\end{equation}
and $D(\sigma\|\tau)=\Tr\sigma\log\sigma-\Tr\sigma\log\tau$ is the quantum relative entropy. The advantage of  formula (\ref{EqD}) is that the right-hand side does not involve the additional register $E$. 

Note that, since $\mathcal Z$ is a partially dephasing channel, Eq.~(\ref{EqD}) is a generalization of a coherence measure proposed in Ref.~\cite{Plenio} called the relative entropy of coherence. Its operational meaning in QKD is investigated in Ref.~\cite{MaNew}. It is the distance between the quantum state $\mathcal G(\rho_{AB})$, emerging as a result of a QKD protocol, and its partially dephased (``partially classical'') counterpart. Thus, the eavesdropper's ignorance in quantum key distribution is equal to a measure of quantumness of the distributed bipartite  state.

The state $\rho_{AB}$ is chosen by Eve and, hence, is unknown to Alice and Bob. So, to make a reliable estimate of the secret key rate, they should minimize  quantity (\ref{EqD}) over all possible quantum states on $\mathcal H_A\otimes\mathcal H_B$ satisfying a set of restrictions. We arrive at the optimization problem:
\begin{subequations}\label{EqRateMax}
\begin{eqnarray}
K&=&\min_{\rho_{AB}\in\mathbf S} D\big(\mathcal G(\rho_{AB})\|\mathcal Z(\mathcal G(\rho_{AB}))\big)-
p_{\rm pass}h(Q_z)\label{EqRateMaxK},\qquad\\
\mathbf S&=&\{\rho\geq0 \text{ on }\mathcal H_A\otimes\mathcal H_B\,\|\Tr\Gamma_i\rho=\gamma_i,\:\forall i\}.\label{EqRateMaxG}
\end{eqnarray}
\end{subequations}
Here the operators $\Gamma_i$ specify the restrictions:
\begin{enumerate}[(1)]
\item Weighted mean detection rate in the $z$ basis:
\begin{equation}\label{EqDetRateW}
\Gamma_{1}=I_2\otimes(\eta P_{z,0}^B+P_{z,1}^B)=p_z\eta I_2\otimes I_2;
\end{equation}
\item Weighted mean error detection rate in the $x$ basis:
\begin{eqnarray}
\Gamma_2&=&\frac1{p_x^2}(P_{x,0}^A\otimes P_{x,1}^B+
\eta P_{x,1}^A\otimes P_{x,0}^B)\nonumber\\&=&
\eta(\ket+\bra+\otimes\ket-\bra-+\ket-\bra-\otimes\ket+\bra+);
\label{EqPhErr}
\end{eqnarray}
\item Fixation of $p_{\rm pass}$, or, in other words, the sifted key rate:
\begin{equation}\label{EqPassFix}
\Gamma_{3}=I_2\otimes (P_{z,0}^B+P_{z,1}^B).
\end{equation}
\end{enumerate}

We have $p_{\rm pass}=p_z\Tr\Gamma_3\rho_{AB}$. Recall that we consider the asymptotic case of infinitely many pulses sent by Alice, $n\to\infty$. In this case, $p_x$ can be made arbitrarily small. The $x$ basis does not participate in the secret key generation; it is used only to estimate  $\gamma_2$. In the limit of infinitely many pulses, an arbitrarily small fraction of them is sufficient to collect a reliable statistics. So, in the following, we put $p_z=1$ and $p_x=0$ in Eq.~(\ref{EqRateMaxK}), but still use the $x$ basis statistics in Eq.~(\ref{EqPhErr}). Then $p_{\rm pass}=\Tr\Gamma_3\rho_{AB}$

Let us discuss the first restriction $\Gamma_1$. Denote  $t=\Tr\rho_{AB}(I_2\otimes I_2)\leq 1$. In the no-eavesdropping case this corresponds to the transparency of the transmission line (so that $1-t$ is the trace of the vacuum component of the Bob's space). Then $\Tr\Gamma_1\rho_{AB}=t\eta$. So, $\Gamma_1$ fixes the transparency for a constant known $\eta$.

Together with the restriction $\Gamma_3$, the restriction $\Gamma_1$ additionally fixes the ratio of zeros and ones in the Bob's sifted key. If the probability of error in the channel in the no-eavesdropping case does not depend on the bit value, then the ratio of the number of ones  to the number of zeros is $\eta$. The restrictions $\Gamma_1$ and $\Gamma_3$ prevent Eve from changing this ratio. As can be seen from Appendix A [see Eqs.~(\ref{EqGammas}) and (\ref{EqGammaRho})], the restriction $\Gamma_1$ can be equivalently replaced by the detection rate of either only zeros or only ones: $\Gamma'_{1}=I_2\otimes P_{z,\beta}^B$, where either $\beta=0$ or $\beta=1$.

Finally, let us denote $Q_x=\Tr\Gamma_2\rho_{AB}/(t\eta)$. In the case of no detection-efficiency mismatch $\eta=1$, this is the QBER in the $x$ basis. Thus,
\begin{subequations}\label{EqRestr}
\begin{eqnarray}
\Tr\Gamma_1\rho_{AB}&=&t\eta,\\
\Tr\Gamma_2\rho_{AB}&=&t\eta Q_x,\\
\Tr\Gamma_3\rho_{AB}&=&p_{\rm pass},
\end{eqnarray}
\end{subequations}
where $t$ and $p_{\rm pass}$ are in the range $(0,1]$ and $Q_x$ is in the range $[0,1]$.

Note also that the error rate in the $z$ basis $Q_z$ in Eq.~(\ref{EqRateMaxK}) is formally given by $Q_z=\Tr\Gamma_4\rho_{AB}/(t\eta)$, where  $\Gamma_4=P_{z,0}^A\otimes P_{z,1}^B+\eta P_{z,1}^A\otimes P_{z,0}^B$. But we do not use this restriction in the optimization problem (\ref{EqRateMax}): Only the first term in the right-hand side of  Eq.~(\ref{EqRateMaxK}) is a subject for optimization and we use only restrictions (\ref{EqDetRateW})--(\ref{EqPassFix}) for it.

\begin{remark}\label{RemRestr}
The ``maximal'' set of restrictions is $\{\Gamma_{jk}=P^A_j\otimes P^B_k\}$ added by the unit trace condition and the fixation of $\rho_A=\Tr_B\rho_{AB}$ by means of the Pauli matrices \cite{Lutk-numeric}. Do some restrictions in addition to (\ref{EqDetRateW})--(\ref{EqPassFix}) allow to obtain  tighter bounds on the Eve's ignorance and, hence, to provide higher secret key rates? The answer depends on the type of the noise in the channel. The natural noise in the transmission line is often described by the depolarizing channel acting in the qubit space:
\begin{equation}\label{EqDep}
\mathcal E_Q(\rho_2)=(1-2Q)\rho_2+2QI_2/2,
\end{equation}
where $Q$ is a QBER in both bases. So, the entangled stated distributed to Alice and Bob in the no-eavesdropping case is:
\begin{eqnarray}
\rho^0_{AB}&=&({\rm Id}_A\otimes\mathcal E_Q)(\ket\Phi_{AB}\bra\Phi)
\oplus0\nonumber\\
&+&(1-t)(I_2/2)\otimes\ket{\rm vac}\bra{\rm vac},\label{EqDepRho}
\end{eqnarray}
where ${\rm Id}_A$ is the identity channel in the Alice's space, $\mathcal E_Q$ acts in the single-photon subspace of the Bob's space, and $\oplus0$ denotes the embedding of the four-dimensional space into the six-dimensional one (with the vacuum component of the Bob's space). Then one can take 
\begin{equation}\label{EqActG}
\gamma_i=\Tr\rho^0_{AB}\Gamma_i
\end{equation}
for all $i$, i.e., $\gamma_1=t\eta$, $\gamma_2=t\eta Q$, and $\gamma_3=t(1+\eta)/2$. In this case, additional restrictions in comparison to Eqs.~(\ref{EqDetRateW})--(\ref{EqPassFix}) does not alter the solution of the optimization problem (\ref{EqRateMax}) and, hence, do not increase the secret key rate; see Remark~\ref{RemTight} in the end of Appendix~A after the proof of the main theorem.

Thus, the bounds for the secret key rate given below are tight whenever the natural noise in the channel is described by the depolarizing channel (\ref{EqDep}) and (\ref{EqDepRho}). In some other cases, additional restrictions may increase the secret key rate \cite{Lutk-num-pre,MaNew,TomogrQKD}.

One may also ask what intuition stands behind a particular choice of the operators $\Gamma_1$, $\Gamma_2$, and $\Gamma_3$. We also postpone the answer to this question until the end of Appendix~A. Here we can say that the matrices of these operators have simple forms; see Eq.~(\ref{EqGammas}). Namely, one can see that, up to a constant factor, the matrix of $\Gamma_2$ coincides with the matrix of the standard (for the case of no detection-efficiency mismatch) phase error rate operator. The value of either $p_{\rm pass}$ or $t=\Tr\rho_{AB}(I_2\otimes I_2)$ is also obviously required to estimate the secret key rate in the case of no efficiency mismatch $\eta=1$. In this case, $p_{\rm pass}=t$. In the case of detection-efficiency mismatch, these are different 
operators and we need both.

\end{remark}

Now we are ready to formulate the main theorem.

\begin{theorem}\label{Th}
Optimization problem (\ref{EqRateMax})--(\ref{EqRestr})  has feasible solutions if and only if 
\begin{equation}\label{EqFeas}
2Q_x\geq1-\sqrt{1-\delta^2},
\end{equation}
where
\begin{equation}\label{EqDelta}
\delta=\frac{2p_{\rm pass}-t(1+\eta)}{t(1-\eta)}.
\end{equation}
In this case, the optimal value of the objective function is given by
\begin{multline}\label{EqMain}
K(Q_z,Q_x,\eta,t,p_{\rm pass})\\=p_{\rm pass}\left[
h\left(\frac{1+\delta}{2p_{\rm pass}}\right)-h(\lambda(Q_x,\eta,t,p_{\rm pass}))-h(Q_z)
\right],
\end{multline}
where
\begin{multline}\label{EqLambda}
\lambda(Q,\eta,t,p_{\rm pass})\\=\frac12-\frac t{4p_{\rm pass}}
\sqrt{[1-\eta+\delta(1+\eta)]^2+4\eta(1-2Q)^2}.
\end{multline}

\end{theorem}

The proof is provided in Appendix~A. Now we consider an important particular case.

\begin{corollary}\label{Cor}
If \begin{equation}\label{EqPpass}
p_{\rm pass}=t(1+\eta)/2,
\end{equation}
then the optimal value of the objective function in optimization problem (\ref{EqRateMax})--(\ref{EqRestr}) is given by
\begin{multline}\label{EqFin}
K(Q_z,Q_x,\eta)\\=p_{\rm pass}\left[
h\left(\frac{1}{1+\eta}\right)-h(\lambda(Q_x,\eta))-h(Q_z)
\right],
\end{multline}
where
\begin{equation}
\lambda(Q,\eta)=\frac12-\frac12
\sqrt{1-\frac{16\eta Q(1-Q)}{(1+\eta)^2}}.
\end{equation}
\end{corollary}

Formulas (\ref{EqMain}) and (\ref{EqFin}) are the desired formulas for the secret key rate. They are tight for the proposed protocol since the optimization problem is solved exactly. 

Condition (\ref{EqFeas}) means that there is no positive semi-definite operator $\rho_{AB}$ satisfying Eq.~(\ref{EqRestr}) if inequality (\ref{EqFeas}) is not satisfied. Hence, the values of $t$, $Q_x$, and $p_{\rm pass}$ not satisfying inequality (\ref{EqFeas}) cannot be obtained.

Let us discuss the difference between formulas (\ref{EqMain}) and (\ref{EqFin}). Suppose that the transmission loss and the probability of error in the channel in the no-eavesdropping case are independent on the bit value (i.e., are the same for the states $\ket 0$ and $\ket 1$). In particular, the depolarizing channel in (\ref{EqDep}) and (\ref{EqDepRho}) satisfies this condition. Then, in the no-eavesdropping case, Eq.~(\ref{EqPpass}) is satisfied: $p_{\rm pass}$ equals  the average between the efficiencies of the detectors multiplied by the transparency $t$. In the asymptotic case $n\to\infty$, Eve cannot violate equality (\ref{EqPpass}) because, otherwise, Alice and Bob will detect the eavesdropping precisely by observing the violation of Eq.~(\ref{EqPpass}). Hence, Eve is restricted to attacks that do not violate Eq.~(\ref{EqPpass}), and the secret key rate is given by Eq.~(\ref{EqFin}).

However, formula (\ref{EqFin}) itself cannot be used in practice because in the finite-key scenario, even in the no-eavesdropping case, the statistical fluctuations lead to deviations of $p_{\rm pass}$ from its mean value $t(1+\eta)/2$. Since we cannot distinguish the statistical deviations from a small $\delta$ introduced by Eve, we must be able to bound the Eve's knowledge on the sifted key for an arbitrary $\delta$ in some neighborhood of zero. Formula (\ref{EqMain}) do this and can be used as a starting point in the finite-key analysis. But in the rest of the paper we will be interested in the actual secret key rate in the asymptotic case and, so, will consider only formula (\ref{EqFin}).

For perfect detection $\eta=1$, formula (\ref{EqFin}) gives the well-known result \cite{ShorPreskill}:
\begin{equation}\label{EqIdeal}
K=t[1-h(Q_x)-h(Q_z)]
\end{equation}
In another particular case of $Q_x=Q_z=0$ (noiseless case),  formula (\ref{EqFin}) gives another known result, which was obtained in Ref.~\cite{Fung}:
\begin{equation}\label{EqNoiseless}
K=p_{\rm pass}h\left(\frac{1}{1+\eta}\right).
\end{equation}

Let us compare formula (\ref{EqFin}) to the two formulas obtained in Ref.~\cite{Fung}:
\begin{eqnarray}
&&K_1=p_{\rm pass}
\left\{
\frac{2\eta}{1+\eta}[1-h(Q_x)]-h(Q_z)
\right\},\label{EqFung1}\\
&&K_2=p_{\rm pass}
\frac{2\eta}{1+\eta}[1-h(Q_z)-h(Q_x)].\label{EqFung2}
\end{eqnarray}
The first formula is obtained as a result of a more general analysis, which includes the cases when $\eta$ is under Eve's control, applied to the particular case of a constant $\eta$. The second formula is obtained by the simple discarding argument. Of course, the simplest solution to the detection-efficiency mismatch problem is to discard every zero from the Bob's raw key (to turn it into the no-click event) with the probability $1-\eta$. This allows us to artificially adjust the efficiencies of the two detectors so that both cases of 0 and 1 bit values are ``detected'' with the probability $\eta$. Thus, the secret key rate is equal to the right-hand side of Eq.~(\ref{EqIdeal}) multiplied by $\eta$, which gives exactly Eq.~(\ref{EqFung2}).

\begin{figure}[t]
\begin{centering}
\includegraphics[width=1\columnwidth]{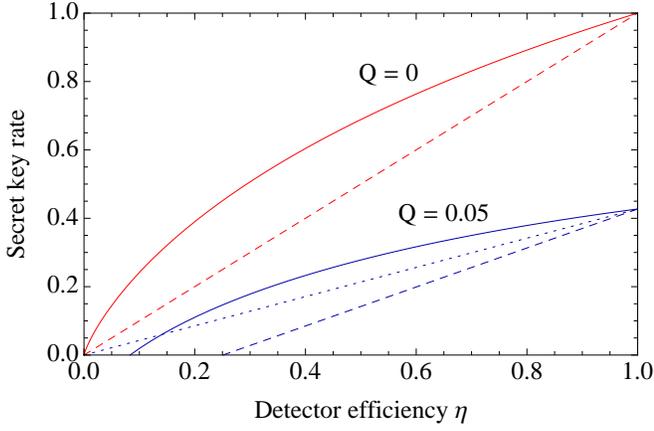}
\end{centering}
\vskip -4mm
\caption
{
Secret key rate $K(Q,Q,\eta)$ of the BB84 protocol vs the efficiency of one of the detectors $\eta$. Another detector and the transmission line  are assumed to be perfect; otherwise, the secret key rate is reduced by a constant factor. Red: the noiseless case $Q=0$. Blue: the case $Q=0.05$. Solid line: formula (\ref{EqFin}) (deviates from the previous one only for the noisy case and small $\eta$). Dashed line: formula (\ref{EqFung1}). Dotted line: formula (\ref{EqFung2}) [coincides with Eq.~(\ref{EqFung1}) for the noiseless case].
}
\label{Fig1}
\end{figure}

The comparison of these formulas are given in Fig.~\ref{Fig1}. We see that formula (\ref{EqFin}) gives definitely higher secret key rates than Eq.~(\ref{EqFung1}). Also it gives higher secret key rates than Eq.~(\ref{EqFung2}) for the most of  range of $\eta$. Only in the case of small $\eta$, which corresponds to a large efficiency mismatch, does the discarding provide higher rates. This does not contradict the tightness of bound (\ref{EqFin}) since the possibility of discarding (a kind of preprocessing of the observation data) immediately after the transmission of quantum states is not provided in our description of the protocol. Also the calculations according to formula (\ref{EqFin}) coincide with the numerical results of Ref.~\cite{Lutk-numeric}.

\begin{figure}[t]
\begin{centering}
\includegraphics[width=1\columnwidth]{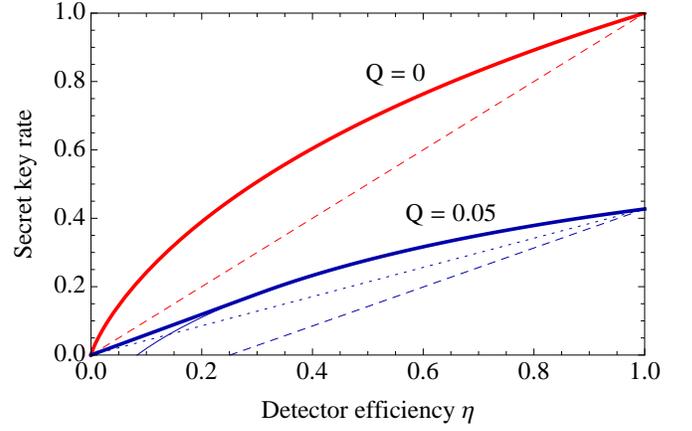}
\end{centering}
\vskip -4mm
\caption
{
Secret key rate $K(Q,Q,\eta)$ of the BB84 protocol vs the efficiency of one of the detectors $\eta$. The lines are the same as in Fig.~\ref{Fig1}, except that the thick solid line has been added. It corresponds to formula (\ref{EqFin2}) and outperforms Eq.~(\ref{EqFin}) (thin solid line) one only for the noisy case and small $\eta$.
}
\label{Fig2}
\end{figure}

\begin{figure}[t]
\begin{centering}
\includegraphics[width=1\columnwidth]{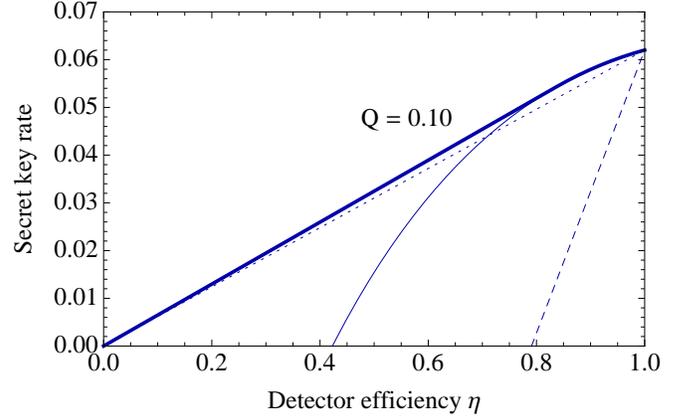}
\end{centering}
\vskip -4mm
\caption
{
Secret key rate $K(Q,Q,\eta)$ of the BB84 protocol vs the efficiency of one of the detectors $\eta$, for $Q=0.10$. The lines are the same as on Fig.~\ref{Fig2}. Formula (\ref{EqFin2})  outperforms Eq.~(\ref{EqFin}) for the most range of $\eta$.
}
\label{Fig3}
\end{figure}

We can modify our protocol to overcome this limitation of formula (\ref{EqFin}). Namely, we can combine the discarding of some zeros (to improve the statistical properties of the Bob's raw key) and calculations according to formula (\ref{EqFin}). Let Bob discards every zero outcome with a probability $1-\eta_1\leq1-\eta$. After that, the ``remaining'' detection-efficiency mismatch corresponds to the  detector efficiency $\eta_2=\eta/\eta_1\geq\eta$ and we can use formula (\ref{EqFin}) with $\eta$ substituted by $\eta_2$. So, we arrive at the formula [for simplicity, let Eq.~(\ref{EqPpass}) be satisfied]
\begin{multline}\label{EqFin2}
K(Q_z,Q_x,\eta)=\max_{
\begin{smallmatrix}
\eta_1\eta_2=\eta,\\
\eta_1\geq\eta.
\end{smallmatrix}
}
\frac{t\eta_1(1+\eta_2)}2\\\times
\left[
h\left(\frac{1}{1+\eta_2}\right)-h(\lambda(Q_x,\eta_2))-h(Q_z)
\right],
\end{multline}
where $t$ and $Q_x$ are related to the observed values of $\Gamma_1$ and $\Gamma_2$ in Eqs.~(\ref{EqDetRateW}) and (\ref{EqPhErr}) with $\eta$ replaced by $\eta_2$. In the limiting cases $(\eta_1=1,\eta_2=\eta)$ and $(\eta_1=\eta,\eta_2=1)$, we obtain formulas (\ref{EqFin}) and (\ref{EqFung2}), respectively. The results of calculations according to  formula (\ref{EqFin2}) are shown in Figs.~\ref{Fig2} and~\ref{Fig3}. We see that Eq.~(\ref{EqFin2}) outperforms Eqs.~(\ref{EqFung1}) and (\ref{EqFung2}) and also outperforms Eq.~(\ref{EqFin}) if the detection-efficiency mismatch or QBER is large.

\begin{figure}[t]
\begin{centering}
\includegraphics[width=1\columnwidth]{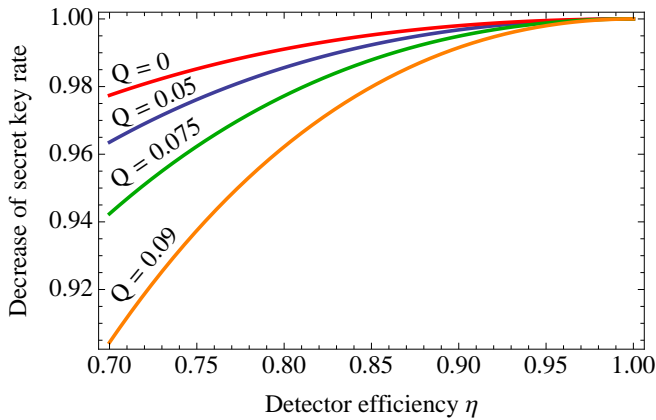}
\end{centering}
\vskip -4mm
\caption
{
Decrease of secret key rate in the detection efficiency-mismatch case with respect to the no-mismatch case: the ratio of the secret key rate in the mismatch case with the detector efficiencies 1 and $\eta$ to the secret key rate in the no-mismatch case with both  efficiencies equal to $(1+\eta)/2$, for various QBERs ($Q_z=Q_x=Q$). If the mismatch is not very large, then the decrease of secret key rate is  relatively small even for high QBERs.
}
\label{Fig4}
\end{figure}

The decrease of secret key rate with the decrease of $\eta$ shown on Figs.~\ref{Fig1}--\ref{Fig3} is caused by two effects: the decrease of the average detector efficiency $(1+\eta)/2$ and detection-efficiency mismatch as such. To distinguish the influence of the mismatch as such, we compare the secret key rates for the mismatch case with the detector efficiencies 1 and $\eta$ and the no-mismatch case with both  efficiencies equal to $(1+\eta)/2$. The secret key rate for the latter case is given by Eq.~(\ref{EqIdeal}) with the right-hand side multiplied by $(1+\eta)/2$. The ratio of the secret key rate in the mismatch case to that in the no-mismatch case for various QBERs is shown on Fig.~\ref{Fig4}. We see that first the secret key rate is larger influenced by mismatch for high QBERs and second, if the mismatch  is not very large, then the decrease of  secret key rate is also relatively small even for high QBERs. For example, the secret key rate for $\eta=0.7$ and $Q=0.09$ is above 90\% of the secret key rate for the no-mismatch case with the same $Q$ and average efficiency.

Finally, let us recall, that, if both detectors are not perfect and have efficiencies $\eta_0$ and $\eta_1$, then $\eta=\min(\eta_0,\eta_1)/\max(\eta_0,\eta_1)$, and the secret key rate  is
\begin{equation}
K(Q_z,Q_x,\eta_0,\eta_1)=\max(\eta_0,\eta_1)K(Q_z,Q_x,\eta)
\end{equation}
with $K(Q_z,Q_x,\eta)$ given by either Eqs.~(\ref{EqMain}), (\ref{EqFin}), or (\ref{EqFin2}).

\section{Decoy state method in the case of detection-efficiency mismatch}\label{SecDecoy}

In this section we adapt the decoy state method to the case of detection-efficiency mismatch. Now we take into account that Alice sends not true single-photon pulses, but weak coherent pulses. We consider the scheme of one signal state and two weak decoy states. This means that each Alice's pulse can be either a signal pulse with the intensity $\mu_{\rm s}=\mu$ (used for key generation) or one of two decoy pulses with the intensities $\mu_{\rm d_1}=\nu_1$ and $\mu_{\rm d_2}=\nu_2$, with the conditions $0\leq\nu_2<\nu_1$ and $\nu_1+\nu_2<\mu$.

We follow the method of Ref.~\cite{MaLo2005}, where a lower bound for the number of detections originated from the single-photon pulses and an upper bound for the error rate for the single-photon pulses were derived. A difference with the case of no efficiency mismatch is that we need separate detection data for each basis and each of Bob's measurement outcome.

At first, let us discuss the scenario  and processing of double clicks. We assume that Alice emits a mixture of $i$-photon states, $i=0,1,2,\ldots$, with the probabilities given by the Poisson distribution with the parameter (the average number of photons) $\mu_v$, $v\in\{{\rm s,d_1,d_2}\}$. Since the outgoing quantum state should be a mixture of Fock states, a passive or active phase randomization  on the Alice's side is required. Eve performs a nondemolition measurement of the number of photons in the pulse, which does not change the state. In the previous sections, Eve was not allowed to send more than one photon to Bob, but now there are multiphoton pulses from the Alice's source. So, we can impose the following restriction instead: Eve is not allowed to add more photons to a pulse. Alternatively, we can impose a weaker restriction: since the  multiphoton states are  treated as insecure anyway, Eve is free to do anything with these states (for example, to add more photons), but she is not allowed to add more photons to the single-photon states. We analyze these assumptions in Remark~\ref{RemDecoyAssump} below.

In this scenario, the double clicks may originate either from the Alice's multiphoton  states or from her single-photon or vacuum states accompanied by dark counts on the Bob's side. Bob may perform a random equiprobable assignment of the double clicks to one of the outcomes \cite{squash}. This  does not affect the security because the Alice's multiphoton  states are anyway treated as insecure and dark counts are not under Eve's control and do not increase her information on the Alice's bit conditioned on the click event. 

Denote by $N^{{\rm s}b}$ the number of positions where both Alice and Bob have chosen the basis $b\in\{z,x\}$ and Alice has sent a signal pulse. Denote by
$M^{{\rm s}b\beta}$ the number of positions where both Alice and Bob have chosen the basis $b$, Alice has sent a signal pulse, and Bob has obtained the result $\beta\in\{0,1\}$.  Denote by $Q^{{\rm s}b\beta}=M^{{\rm s}b\beta}/N^{{\rm s}b}$ the gain for the signal states, the basis $b$, and the outcome $\beta$. 

Correspondingly, denote by $Q^{{\rm d_1}b\beta}$ and $Q^{{\rm d_2}b\beta}$ the gains for the two types of decoy states, the basis $b$, and the outcome $\beta$. These gains should not be confused with the QBERs $Q_x$ and $Q_z$ from the previous sections: In this section, we use the notations that are common for the literature on the decoy state method. For QBERs, we will use other notations instead, see below. 

Denote also by $Y_i^{b\beta}$ the yield of an $i$-photon state, the basis $b$, and the Bob's outcome $\beta$, i.e., the conditional probability of a detection event with the outcome $\beta$ given that Alice sends out an $i$-photon state and both Alice and Bob choose the basis $b$. The yields $Y_i^{b\beta}$ for all $i$ are assumed to be under Eve's control. More precisely, in the finite-key analysis, $Y_i^{b\beta}$ is defined \textit{a posteriori} \cite{Ma2017}: as a ratio of the number of $i$-photon states (with the specified $b$ and $\beta$) received by Bob  to the number of such states sent by Alice. But the numerator in this fraction is under Eve's control. However, a crucial trick lying in the heart of the decoy state method is that Eve observes the number of photons in a pulse  but cannot obtain knowledge on the intensity (the parameter of the Poisson distribution) of the pulse. Alice announces the intensity of each pulse ($\mu,\nu_1$, or $\nu_2$) on the post-processing stage. Processing of the detection data separately for each type of pulses leads to detection of the PNS attack and, more generally, allows us to estimate the number of single-photon (i.e., secure) states received by Bob.

Note that the yield of an $i$-photon state depends on the basis in the case of efficiency mismatch. For example, the incoming Bob's single-photon state $\ket 0$ will be definitely detected if it is measured in the $z$ basis, but it will be detected with the probability $(1+\eta)/2$ if it is measured in the $x$ basis. In Ref.~\cite{DecoyMismatch}, an additional implicit assumption is imposed that $Y_i$ does not depend on the basis. This is a restriction on the class of possible Bob's incoming states and, hence, on the class of manipulations available for Eve. Our analysis does not require such a restriction.

We have 
\begin{equation}\label{EqGainSum}
Q^{vb\beta}=
\sum_{i=0}^\infty Y_i^{b\beta}\frac{\mu_v^i}{i!}e^{-\mu_v}
=\sum_{i=0}^\infty Q_i^{vb\beta},
\end{equation}
$v\in\{\rm s,d_1,d_2\}$, where 
\begin{equation}\label{EqQi} 
Q_i^{vb\beta}=Y_i^{b\beta}\frac{\mu_v^i}{i!}e^{-\mu_v}
\end{equation}
is the gain for  $i$-photon states of the type $v\in\{\rm s,d_1,d_2\}$, the basis $b$, and the outcome $\beta$. Using  formula (\ref{EqGainSum}), we can repeat the derivation in Ref.~\cite{MaLo2005} to obtain a lower bound for the number of detections with the outcome $\beta$ in the $z$ basis originated from the single-photon signal pulses:

\begin{multline}\label{EqQ1L}
Q_1^{{\rm s}z\beta}\geq Q_1^{{\rm L s}z\beta}=
\frac{\mu^2e^{-\mu}}{\mu\nu_1-\mu\nu_2-\nu_1^2+\nu_2^2}
\\
\times
[Q^{{\rm d_1}z\beta}e^{\nu_1}-Q^{{\rm d_2}z\beta}e^{\nu_2}
-\frac{\nu_1^2-\nu_2^2}{\mu^2}(Q^{{\rm s}z\beta}e^{\mu}-Y_0^{{\rm L}z\beta})],
\end{multline}
where 
\begin{equation}\label{EqY0L}
Y_0^{{\rm L}z\beta}=
\max\{\frac{
\nu_1Q^{{\rm d_2}z\beta}e^{\nu_2}
-
\nu_2Q^{{\rm d_1}z\beta}e^{\nu_1}
}{\nu_1-\nu_2},0\}.
\end{equation}
The upper bound is trivial: 
$Q_1^{{\rm s}z\beta}\leq Q^{{\rm s}z\beta}$. Note that the conditions $\nu_2<\nu_1$ and $\nu_1+\nu_2<\mu$ (see the beginning of this section) are used in the derivation of estimate (\ref{EqY0L}).

Denote by
$M^{vb\beta}_{\rm err}$ the number of positions where both Alice and Bob have chosen the basis $b$, Alice has sent a pulse of the type $v$ encoding the bit value $1-\beta$, and Bob has obtained the result $\beta$. Denote by $E^{vb\beta}=M^{vb\beta}_{\rm err}/M^{vb\beta}$ the QBER for the pulses of the type $v$, the basis $b\in\{x,z\}$, and the Bob's outcome $\beta$. If we denote
\begin{equation}
Q^{vb}=Q^{vb0}+Q^{vb1}
\end{equation}
as the total gain for the $v$-pulses and the $b$ basis, and 
\begin{equation}
E^{vb}=\frac{M^{vb0}_{\rm err}+M^{vb1}_{\rm err}}{M^{vb0}+M^{vb1}}
\end{equation}
as the usual QBER in the $b$ basis, then
\begin{equation}
E^{vb}Q^{vb}=E^{vb0}Q^{vb0}+E^{vb1}Q^{vb1}.
\end{equation}
Analogously, denote by $e_i^{b\beta}$ the the QBER for the $i$-photon states, the basis $b$, and the outcome $\beta$. Like $Y_i^{b\beta}$, $e_i^{b\beta}$ is also assumed to be under the Eve's control. We have

\begin{equation}\label{EqEQsum}
E^{vb\beta}Q^{vb\beta}=
\sum_{i=0}^\infty e_i^{b\beta}Q_i^{vb\beta},
\end{equation}
\begin{equation}\label{EqEboundpre}
E^{{\rm d_1}x\beta}Q^{{\rm d_1}x\beta}e^{\nu_1}
-
E^{{\rm d_2}x\beta}Q^{{\rm d_2}x\beta}e^{\nu_2}
\geq e_1^{x\beta} Y_1^{x\beta}(\nu_1-\nu_2),
\end{equation}
\begin{equation}\label{EqEbound}
e_1^{x\beta}Q_1^{{\rm s}x\beta}\leq
\left(E^{{\rm d_1}x\beta}Q^{{\rm d_1}x\beta}e^{\nu_1}
-
E^{{\rm d_2}x\beta}Q^{{\rm d_2}x\beta}e^{\nu_2}
\right)
\frac{\mu e^{-\mu}}{\nu_1-\nu_2}.
\end{equation}

Now we apply these results to modify formula (\ref{EqMain}) for weak coherent pulses. The last term $p_{\rm pass}h(Q_z)$ corresponds to the error correction in all detected signal bits. Hence, it should be rewritten as $Q^{{\rm s}z}h(E^{{\rm s}z})$. 

The first two terms in Eq.~(\ref{EqMain}) correspond to the Eve's ignorance and, hence, should obtain only single-photon contributions: Information encoded in multiphoton pulses is assumed to be completely known to Eve. So, let $\rho_{AB}$ be the (unnormalized) density operator containing only the vacuum and single-photon parts of signal pulses. Then, according to the previous definitions,
\begin{equation}
\Tr (I_2\otimes P^B_{b\beta})\rho_{AB}=Q_1^{{\rm s}b\beta},
\end{equation}
and hence
\begin{equation}\label{Eqtpdecoy}
t=Q_1^{{\rm s}z0}+Q_1^{{\rm s}z1}/\eta,\qquad
p_{\rm pass}=Q_1^{{\rm s}z0}+Q_1^{{\rm s}z1},
\end{equation}
and
\begin{equation}\label{Eqq}
\Tr\Gamma_2\rho_{AB}\equiv\gamma_2=
\eta e_1^{{\rm s}x0}Q_1^{{\rm s}x0}+
e_1^{{\rm s}x1}Q_1^{{\rm s}x1}.
\end{equation}
So, the first two terms in Eq.~(\ref{EqMain}) should be modified to
\begin{equation}
p_{\rm pass}\left[
h\left(
\frac{1+\delta}{2p_{\rm pass}}\right)
-
h(\lambda(\gamma_2/(t\eta),
\eta,t,p_{\rm pass}))
\right]
\end{equation}
and the secret key rate is equal to
\begin{multline}\label{EqMainDecoyPre}
K(Q_1^{{\rm s}z0},Q_1^{{\rm s}z1},\gamma_2)\\=
p_{\rm pass}\left[h\left(
\frac{1+\delta}{2p_{\rm pass}}\right)
\!-
h(\lambda(\gamma_2/(t\eta),
\eta,t,p_{\rm pass}))
\right]\\
-Q^{{\rm s}z}h(E^{{\rm s}z}),
\end{multline}
where $\delta$ is, as before, defined by Eq.~(\ref{EqDelta}) and $t$, $p_{\rm pass}$, and $\gamma_2$ are defined by Eqs.~(\ref{Eqtpdecoy}) and (\ref{Eqq}). 

But now, in contrast to the previous sections, the quantities $Q_1^{{\rm s}z1},Q_1^{{\rm s}z1}$, and $\gamma_2$ are not given directly; we have only estimates. Hence, we should consider the worst-case scenario and  take values of these parameters within the estimated ranges that give the minimal value to  function (\ref{EqMainDecoyPre}). 

Function (\ref{EqMainDecoyPre}) monotonically decreases with the increase of $\gamma_2$. This can be established by a mathematical investigation of this function and also is obvious from the physical meaning: The secret key rate certainly decreases with the increase of the error rate. Hence, the worst-case scenario corresponds to the maximally possible value of $\gamma_2$. In other words, we should use an upper bound for $\gamma_2$. Using estimate (\ref{EqEbound}), we obtain $\gamma_2\leq q\eta$, where 
\begin{multline}
q=\Big[
\left(E^{{\rm d_1}x0}Q^{{\rm d_1}x0}+
E^{{\rm d_1}x1}Q^{{\rm d_1}x1}/\eta\right)e^{\nu_1}
\\
-
\left(E^{{\rm d_2}x0}Q^{{\rm d_2}x0}+
E^{{\rm d_2}x1}Q^{{\rm d_2}x1}/\eta\right)
e^{\nu_2}
\Big]
\frac{\mu e^{-\mu}}{\nu_1-\nu_2}.
\end{multline}

The dependence of  function (\ref{EqMainDecoyPre}) on $Q_1^{{\rm s}z0}$ and $Q_1^{{\rm s}z1}$ is, in general, nonmonotonic, so, we should minimize over these quantities in the range 
\begin{equation}\label{EqQRange}
Q_1^{{\rm Ls}z\beta}\leq Q_1^{{\rm s}z\beta}\leq Q^{{\rm s}z\beta},\quad \beta=0,1,
\end{equation}
with $Q_1^{{\rm Ls}z\beta}$ defined in Eq.~(\ref{EqQ1L}). 

However, in the simulation on Fig.~\ref{Fig5} below, the minimum is always achieved in the lower bounds: $Q_1^{{\rm s}z\beta}=Q_1^{{\rm Ls}z\beta}$, $\beta=0,1$. The reason is that function (\ref{EqMainDecoyPre}) monotonically decreases with the decrease of $Q_1^{{\rm s}z0}$ or $Q_1^{{\rm s}z1}$ in a neighborhood of a point $(\bar Q_1^{{\rm s}z0},\bar Q_1^{{\rm s}z1})$ if $\bar Q_1^{{\rm s}z1}/\bar Q_1^{{\rm s}z0}=\eta$. Such points correspond to the case of the depolarizing channel (\ref{EqDepRho}) and equality (\ref{EqPpass}). Hence, if the lower bounds $Q_1^{{\rm Ls}z\beta}$ are not too far from the actual values of $Q_1^{{\rm s}z\beta}$ and belong to this neighborhood, the minimum is achieved in these lower bounds. However, for generality, we leave the minimization operation.

Thus, the final formula for the secret key rate is
\begin{multline}\label{EqMainDecoy}
K=\min_{Q_1^{{\rm s}z0},Q_1^{{\rm s}z1}}
p_{\rm pass}\left[h\left(
\frac{1+\delta}{2p_{\rm pass}}\right)
\!-
h(\lambda(q/t,
\eta,t,p_{\rm pass}))
\right]\\
-Q^{{\rm s}z}h(E^{{\rm s}z}),
\end{multline}
where the minimization is over range (\ref{EqQRange}).

Finally, let us note that if we express $\delta$ directly through $Q_1^{{\rm s}z0}$ and $Q_1^{{\rm s}z1}$, then Eq.~(\ref{EqMainDecoy}) can be rewritten in the following way, which is, in some sense, simpler and more intuitive:
\begin{multline}\label{EqMainDecoy2}
K=\min_{Q_1^{{\rm s}z0},Q_1^{{\rm s}z1}}
p_{\rm pass}\left[
h(\widetilde\lambda(t/2,Q_1^{{\rm s}z0},Q_1^{{\rm s}z1},
\eta))
\right.\\\left.
-
h(\widetilde\lambda(q,Q_1^{{\rm s}z0},Q_1^{{\rm s}z1},
\eta))
\right]
-Q^{{\rm s}z}h(E^{{\rm s}z}),
\end{multline}
where
\begin{multline}
\widetilde\lambda(q,Q_1^{{\rm s}z0},Q_1^{{\rm s}z1},
\eta)\\=
\frac12-\frac 1{2p_{\rm pass}}
\sqrt{\left(Q_1^{{\rm s}z0}-Q_1^{{\rm s}z1}\right)^2+\eta(t-2q)^2}.
\end{multline}

\begin{remark}\label{RemDecoyAssump}
Now let us discuss the role of the assumption (see the beginning of this section) that Eve is free to do anything with multiphoton states (for example, to add more photons), but is not allowed to add more photons to the single-photon states. The analysis of detection data in the decoy state BB84 protocol consists of two parts: 
\begin{enumerate}[(i)]
\item estimating the number of single-photon signal pulses and QBERs for  single-photon states, and

\item estimating the Eve's information about the part of the sifted key originated from  single-photon states. This part uses the results of part (i) and the results of a security analysis for the case of a true single-photon source.
\end{enumerate}
Information encoded in  multiphoton states is assumed to be completely known to Eve.  

In other words, such a scheme allows us to adapt the results obtained for  a true single-photon source to the case of weak coherent pulses. But the assumptions made in the analysis for weak coherent pulses must be in agreement with those made in the analysis for single-photon states.

In our analysis  for single-photon states we assumed that Eve is not allowed to add more photons to such pulses. Hence, in order to use the results of this analysis [namely, formula (\ref{EqMain})]  in part (ii), we should assume that if a pulse contains only one photon, Eve cannot add more to it.

Now let us consider part (i). A  nice feature of the estimations of the fractions of single-photon states and the corresponding QBERs developed in Ref.~\cite{MaLo2005} is that they actually do not impose any assumptions on the efficiency of detectors  or on  particular forms of $Y_i^{b\beta}$ and $e_i^{b\beta}$.  Inequalities (\ref{EqQ1L}) and (\ref{EqEbound}) are based only on  non-negativity of $Y_i^{b\beta}$ and $e_i^{b\beta}$. Namely, inequality (\ref{EqEboundpre}) is a simple  algebraic consequence of Eqs.~(\ref{EqEQsum}) and (\ref{EqQi}) together with non-negativity of $Y_i^{b\beta}$ and $e_i^{b\beta}$. The derivation of estimate (\ref{EqQ1L}) is a bit more complicated and is given in Ref.~\cite{MaLo2005} but also is purely algebraic and based solely on non-negativity of $Y_i^{b\beta}$ and $e_i^{b\beta}$. This actually means that estimates (\ref{EqQ1L}) and (\ref{EqEbound}) are still valid even if Eve is free to set $Y_i^{b\beta}$ and $e_i^{b\beta}$ for all $i$ to arbitrary values. For example, she is free to add more photons to $i$-photon states to increase $Y_i^{b\beta}$, etc. In principle,  solely in part (i), we can even allow Eve to have full control on the detector efficiencies.

Thus, we need not have any assumptions on the Eve's abilities at all for part (i), i.e., for inequalities (\ref{EqQ1L}) and (\ref{EqEbound}). We only need to forbid Eve to add more photons to single-photon states  in order to use the results of the analysis for a true single-photon source in part (ii). 

The separation of the detection data for each basis and each measurement outcome is also needed for part (ii) but not part (i). Estimation of the aggregated quantities $Q_1^{\rm s}$ and $e_1$ is also possible in part (i) because, again, no assumption except non-negativity of yields and error rates is used in the estimations in Ref.~\cite{MaLo2005}.  But  quantities (\ref{Eqtpdecoy}) and (\ref{Eqq}), which are necessary for Eq.~(\ref{EqMain}), require separate data for each basis and each outcome (the latter is because  $Q_1^{{\rm s}b0}$ and  $Q_1^{{\rm s}b1}$ enter these quantities with different weights).
\end{remark}

The results of calculations of the secret key rate for the decoy state protocol according to formula (\ref{EqMainDecoy2}) [or (\ref{EqMainDecoy})] is given on Fig.~\ref{Fig5}. The parameters have been chosen as follows: the intensity of the signal state $\mu=0.5$, the intensities of two decoy states  $\nu_1=0.1$ and $\nu_2=0$, the fiber attenuation coefficient $\delta=0.2$~dB/km, additional losses in the Bob's optical scheme $t_{\rm Bob}=5$~dB, the optical error probability (see Ref.~\cite{MaLo2005}) $e_{\rm det}=0.01$ , the efficiencies of the detectors $\eta_0=0.1$ and $\eta_1=0.07$ (i.e., $\eta=\eta_1/\eta_0=0.7$),  and the dark count probability per pulse for each detector $Y_0^{\beta=0}=Y_0^{\beta=1}=10^{-6}$. 

We compare the secret key rate according to formulas (\ref{EqMainDecoy}) and (\ref{EqMainDecoy2}) with a theoretical limit: formula (\ref{EqMainDecoyPre}) with the actual values of $Q_1^{{\rm s}z1},Q_1^{{\rm s}z1}$, and $\gamma_2=\Tr\Gamma_2\rho_{AB}$  (given by Eq.~(\ref{Eqq})). 
 
 For the calculation of the actual values  of these quantities, we employ the standard model of losses and errors in a fiber-based QKD setup; see, e.g., Ref.~\cite{MaLo2005}. The probability that a photon emitted by Alice will reach the Bob's detectors is $10^{-(\delta l+\delta_{\rm Bob})/10}$, where $l$ is the transmission distance in kilometers.  Since this probability   is rather small even for $l=0$ and is very small for realistic distances, the probability that Alice's $i$-photon state will reach  Bob's detectors can be approximately taken as $i10^{-(\delta l+\delta_{\rm Bob})/10}$. This approximation actually means that we neglect the possibility that more than one photon from  Alice's pulse will reach  Bob's detectors. 
 
The optical error probability $e_{\rm det}$ is the probability that a photon that has reached  Bob's detectors will hit the wrong detector. Then the detection in the detector $\beta\in\{0,1\}$ occurs if either Alice sends the bit $\beta$ (with the probability 1/2) and no optical error occurs (with the probability $1-e_{\rm det}$)  or Alice sends the bit $1-\beta$ (also with the probability 1/2) and an optical error occurs (with the probability $e_{\rm det}$). In both cases the pulse should reach the detectors (with the probability $i10^{-(\delta l+\delta_{\rm Bob})/10}$) and the detector $\beta$ should register the photon (with the probability $\eta_\beta$). One more possibility is the dark count of the detector $\beta$, which occurs with the probability $Y^\beta_0$. Again, we neglect the probability of simultaneous dark count with the detection of a photon since both probabilities are small. Thus, the actual value of $Y_i^{b\beta}$ is given by
\begin{eqnarray}
Y_i^{b\beta}&=&Y^\beta_0+
i10^{-\frac{\delta l+\delta_{\rm Bob}}{10}}
\frac{(1-e_{\rm det})+
e_{\rm det}}2\eta_{\beta}\nonumber\\
&=&Y^\beta_0+i10^{-\frac{\delta l+\delta_{\rm Bob}}{10}}\eta_{\beta}/2.
\label{EqYiActual}
\end{eqnarray} 
The actual values of $Q^{vb\beta}$ and  $Q_1^{vb\beta}$ are given then by Eqs.~(\ref{EqGainSum}), (\ref{EqQi}), and (\ref{EqYiActual}). The actual values of $e_i^{b\beta}$ are given by
\begin{equation}\label{Eqeactual}
e_i^{b\beta}=
\frac{Y^\beta_0+
i10^{-\frac{\delta l+\delta_{\rm Bob}}{10}}
e_{\rm det}\eta_\beta}
{2Y_i^{b\beta}},
\end{equation}
where we have taken into account that a dark count can be correct or erroneous with the probability 1/2. Then the actual values of $E^{vb\beta}$ and $\gamma_2=\Tr\Gamma_2\rho_{AB}$ are given by  Eqs.~(\ref{EqEQsum}), (\ref{Eqq}), and (\ref{Eqeactual}). Note that both $Y_i^{b\beta}$ and $e_i^{b\beta}$ are basis-independent in this model.

A comparison of the secret key rate with the described theoretical limit shows the quality of estimates (\ref{EqQ1L}) and (\ref{EqEbound}). As we see from Fig.~\ref{Fig5}, the secret key rate is very close to the theoretical limit. This is not surprising since, as is shown in Ref.~\cite{MaLo2005}, estimates (\ref{EqQ1L}) and (\ref{EqEbound}) are tight in the limit $\nu_1\to0$, $\nu_2=0$. 

Also, we compare the secret key rate with the theoretical limit in the case of no efficiency-mismatch but the same average detection efficiency $(\eta_0+\eta_1)/2$ (like on Fig.~\ref{Fig4}). This comparison shows how much the secret key rate in the decoy state protocol is decreased by detection-efficiency mismatch. In agreement with Fig.~\ref{Fig4}, we see that the decrease is rather small.

\begin{figure}[t]
\begin{centering}
\includegraphics[width=1\columnwidth]{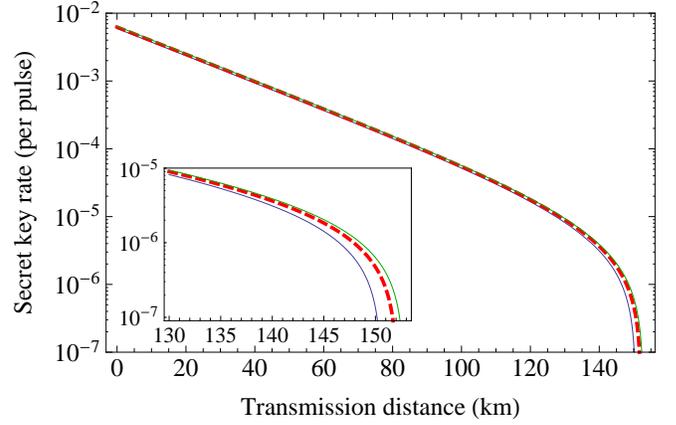}
\end{centering}
\vskip -4mm
\caption
{
Secret key rate of the decoy state BB84 protocol with detection-efficiency mismatch. The parameters are as follows: the intensity of the signal state, $\mu=0.5$; the intensities of and two decoy states,  $\nu_1=0.1$ and $\nu_2=0$; the fiber attenuation coefficient, 0.2~dB/km; additional losses in the Bob's optical scheme, 5~dB; the optical error probability $e_{\rm det}=0.01$; the efficiencies of the detectors, $\eta_0=0.1$ and $\eta_1=0.07$ (i.e., $\eta=\eta_1/\eta_0=0.7$); and  the dark count probability per pulse for each detector, $Y_0^{\beta=0}=Y_0^{\beta=1}=10^{-6}$. Blue line: performance of the protocol according to derived formula (\ref{EqMainDecoy2}) [or (\ref{EqMainDecoy})]. Red dashed line: theoretical limit, formula (\ref{EqMainDecoyPre}) with the actual values of $Q_1^{{\rm s}z1},Q_1^{{\rm s}z1}$, and $\gamma_2=\Tr\Gamma_2\rho_{AB}$ [given by Eq.~(\ref{Eqq})]. Green line: theoretical limit for  the case of no efficiency mismatch but the same average detection efficiency $(\eta_0+\eta_1)/2$ (like on Fig.~\ref{Fig4}).
}
\label{Fig5}
\end{figure}

\section{Discussions and conclusions}
We have proved the security of the BB84 QKD protocol with  detection-efficiency mismatch under the assumption that the eavesdropper sends no more than one photon to the legitimate receiver if the legitimate resender's pulse is single photon. We have derived tight bounds on the secret key rate: formulas (\ref{EqMain}), (\ref{EqFin}) and (\ref{EqFin2}). We used the approach of Refs.~\cite{Lutk-numeric,Lutk-num-pre} based on a reduction of the determination of the secret key rate to a convex optimization problem: the minimization of the relative entropy of coherence.  We have demonstrated that the analytical (rather than numerical) minimization of the relative entropy of coherence also can be used as a method of solving  QKD problems. 

Also we have proposed an adaptation of the decoy state method to the case of efficiency mismatch and obtained formulas (\ref{EqMainDecoy}) and (\ref{EqMainDecoy2}) for the secret key rate in the case of weak coherent pulses on  Alice's side.

The knowledge of an analytic expression for the secret key rate has certain advantages against a numerical result: The former simplifies the calculation of the secret key rate and the analysis of its dependence on the parameters and observations ($\gamma_i$). In particular, the knowledge of an analytic expression allowed us to adapt the decoy state method in a rather simple way. Also an analytic derivation provides deeper understanding of the corresponding scenario. Finally, the presented analytic approach, in principle, can deal with multidimensional and even infinite-dimensional systems; see below. From the other side, in complicated scenarios, the numerical approach allows us to obtain more tight bounds for the secret key rate. So, numerical and analytic approaches to the minimization of the relative entropy of coherence complement each other.

A finite-key analysis can be developed starting from formula (\ref{EqMain}) by the use of conservative (pessimistic) statistical estimations of parameters and the entropy accumulation technique \cite{EntrAccum}. Also recall that the Devetak and Winter formula for the secret key rate (\ref{EqDW}) is valid only for collective Eve's attacks. Using the entropy accumulation technique, the security against the most general, coherent attacks can be proved.

The main remaining problem is the proof of the security of QKD with  detection-efficiency mismatch for  the case when Eve is allowed to send an arbitrary number of photons to Bob. She can do this, for example, to artificially increase the efficiency of one of the detectors. Also the analysis of the case when the efficiencies of the detectors are  under partial Eve's control is important due to experimental realizations of such attacks \cite{LoHack,Makarov}.

These open problems naturally raise the question whether the presented  approach based on the analytic minimization of the relative entropy of coherence can deal with multidimensional or even infinite-dimensional systems. The removal of either the single-photon assumption or the constant efficiency mismatch leads to the increase of the dimensionality. Of course, the direct solution of eigenvalue problems, which was used in the proof of Theorem~\ref{Th}, is impossible in the  multidimensional case. However, a possible way to deal with this case is the use of some quantum channel $\Phi$ which reduces the dimensionality. Since the quantum relative entropy cannot increase under the action of a quantum channel simultaneously on both arguments, we have 
\begin{multline}\label{EqReduc}
D\big(\Phi(\mathcal G(\rho_{AB}))\|\Phi(\mathcal Z(\mathcal G(\rho_{AB})))\big)\\
\leq D\big(\mathcal G(\rho_{AB})\|\mathcal Z(\mathcal G(\rho_{AB}))\big).
\end{multline}
Hence, the minimization of the low-dimensional left-hand side of Eq.~(\ref{EqReduc}) yields the lower bound on the secret key rate. A difficulty is that the channel $\Phi$ is applied after the post-processing, but the constraints are on $\rho_{AB}$. So, the optimization problem is still applied to the high-dimensional space. However, since $\Phi$ somehow aggregates the high-dimensional data, we can hope that it could lead to certain simplification of the problem.

Thus, to extend the results of this paper to the multiphoton case, various techniques can be used. The first way is the use of the detector decoy method \cite{DetDecoy}. The second way is to bound the number of the multiphoton contributions on the Bob's side using the techniques from Ref.~\cite{LutkEnt} based on the monitoring of the average error rate for both bases and the average double click rate for both bases. The third way is a reduction to a low-dimensional case and the use of formula (\ref{EqReduc}) to bound the Eve's information originated from the multiphoton contributions on the Bob's side. Since the detectors are still ``binary'' (click or no click) and cannot capture the detailed structure of multidimensional states, we can hope that an appropriate reduction to a low-dimensional case might exist. A reduction channel $\Phi$ might be in some sense similar to quantum channels using in  squashing models \cite{squash}. But an appropriate channel $\Phi$ might exist even if a squashing model does not exist because we do not require any kind of equivalence relation between the high- and low-dimensional systems. We only need is to find a channel which reduces the dimensionality, but does not give too much advantage to Eve. Also one can try to combine the numerical minimization of the Eve's ignorance on a subspace corresponding to a restricted number of photons with an analytic estimation of the Eve's ignorance on the complement (infinite-dimensional) subspace by means of a reduction $\Phi$ and the minimization of the left-hand side of Eq.~(\ref{EqReduc}) using the techniques proposed here.

Preliminary results of this paper (including formula (\ref{EqFin})) were presented at the International conference "Quantum information, statistics, probability" with a special session dedicated to A. S. Holevo's 75-th birthday \cite{Trush}.

\begin{remark}
Very recently, during the preparation of this paper, another paper \cite{MaNew} was published with the derivation of formula (\ref{EqFin}) by a similar method (also based on the analytical minimization of the relative entropy of coherence). The differences of our result are as follows. First, in Ref.~\cite{MaNew}, a closed analytic formula (\ref{EqFin}) is derived under an additional assumption that  Eve's attack is symmetric. In contrast, we prove the validity of Eq.~(\ref{EqFin}) for arbitrary Eve's collective attacks, but with a weaker assumption (\ref{EqPpass}) on the transmission line in the no-eavesdropping case. Formula (\ref{EqMain}) is valid without this assumption and can be used as a starting point for a finite-key analysis. Second, we additionally analyze the case when Eve can use the vacuum component of the Bob's space. We prove that this  does not give an advantage to Eve, but, \textit{a priori}, this was not obvious in the case of detection-efficiency mismatch (see the discussion in the beginning of Sec.~\ref{SecPM}). Third, we have derived a slightly modified formula (\ref{EqFin2}), which outperforms Eq.~(\ref{EqFin}) if the detection-efficiency mismatch or QBER is large.
\end{remark}

{\bf Acknowledgments}.
We are grateful to A.\,V.\,Duplinskiy, A.\,K.\,Fedorov, 
N.\,L\"utkenhaus, X.\,Ma, D.\,V.\,Sych, I.\,V.\,Volovich, and Y.\,Zhou for fruitful discussions and comments. This work was supported by the Russian Science Foundation (Project No.~17-11-01388).

\setcounter{equation}{0}
\setcounter{section}{0}
\renewcommand{\theequation}{A\arabic{equation}}

\section*{Appendix A. Proof of Theorem \ref{Th}}

\subsection{Preliminaries} 
Define
\begin{equation}\label{EqTarget}
D\Big(\mathcal G(\rho_{AB})\|\mathcal Z\big(\mathcal G(\rho_{AB})\big)\Big)=f(\rho_{AB}).
\end{equation}
Sometimes we will omit the subindexes $AB$ of $\rho_{AB}$. The gradient of this function is given by \cite{Lutk-numeric}
\begin{equation}\label{EqGrad}
\nabla f(\rho)=\mathcal G^\dag\Big(\log\mathcal G(\rho)- 
\log\mathcal Z\big(\mathcal G(\rho)\big)\Big)^T,
\end{equation}
where $\mathcal G^\dag$ is the dual quantum channel to $\mathcal G$. Since $\rho$ can be specified by a finite number of real numbers, the gradient can be understood as the usual gradient of a function of several real variables.

If the projections of the gradient (\ref{EqGrad}) to all allowable directions of movement away from $\rho$ (i.e., the directions that do not violate the positivity of $\rho$ and all restrictions) are non-negative, then $\rho$ yields a local minimum to the objective function (\ref{EqTarget}). Because of the well-known property of the joint convexity of the quantum relative entropy \cite{Lieb} and due to the convexity of the set $\mathbf S$ in Eq.~(\ref{EqRateMax}), a local minimum is a global one. Hence, the non-negativity of the projections of the gradient to all allowable directions is a sufficient condition for the global minimum of the secret key rate.

However, this is not a necessary condition since the gradient (\ref{EqGrad}) may be ill defined. First, if $\rho$ belongs to the boundary of the set $\mathbf S$, then the gradient is ill-defined by definition. Second, the operator $\mathcal G(\rho_{AB})$ is typically degenerate. This is not a problem for the relative entropy expression (\ref{EqD}) due to the rule $0\log0=0$ and the fact that the kernel subspace of $\mathcal  Z(\mathcal G(\rho))$ is a subspace of the kernel subspace of $\mathcal G(\rho)$. But the degeneracy may be a problem for the gradient expression (\ref{EqGrad}) if the kernel subspace of $\mathcal G(\rho)$ is not a subspace of the kernel subspace of  $\mathcal G(\rho+\Delta\rho)$ for an arbitrarily small $\Delta\rho$. In this case, the gradient may be infinite (since the derivative of $x\log x$ is infinite for $x=0$).

Let us consider this problem. The general form of $\rho_{AB}$ is
\begin{equation}
\rho_{AB}=\sum_{i,k\in\{0,1\}}\sum_{j,l\in\{0,1,{\rm vac}\}}
\rho_{ij,kl}\ket{ij}\bra{kl}.
\end{equation}
Let us show that, if the operator
\begin{equation}
\rho'_{AB}=\sum_{i,j,k,l=0}^1
\rho_{ij,kl}\ket{ij}\bra{kl}
\end{equation}
is non-degenerate, then the gradient (\ref{EqGrad}) is well defined by continuity. We have
\begin{eqnarray}
\mathcal G(\rho_{AB})&&=\ket{zz}_{\widetilde A\widetilde B}\bra{zz}
\nonumber
\\&&\otimes \sum_{i,j,k,l=0}^1\eta^{(j+l)/2}\rho_{ij,kl}
\nonumber
\ket{ij}_{AB}\bra{kl}
\otimes
\ket{ij}_{\overline A\,\overline B}\bra{kl}.
\end{eqnarray}
We see that $\mathcal G(\rho_{AB})$ is degenerate on the space $A\widetilde A\overline AB\widetilde B\overline B$. Namely,
\begin{equation}
\mathcal G(\rho_{AB})\ket{ij}_{AB}\ket{i'j'}_{\overline A\,\overline B}
\ket{ab}_{\widetilde A\widetilde B}=0
\end{equation}
whenever $i\neq i'$, $j\neq j'$, $a\neq z$, or $b\neq z$.
Denote by $P_0$ the projector onto the kernel subspace of $\mathcal G(\rho_{AB})$. The dual channel $\mathcal G^\dag$ acts on an arbitrary operator $C$ in the space $A\widetilde A\overline AB\widetilde B\overline B$ as:
\begin{equation}\label{EqGdag}
\mathcal G^\dag(C)=
\sum_{i,j,k,l=0}^1\eta^{(j+l)/2}
c_{ij,kl}
\ket{ij}_{AB}\bra{kl},
\end{equation}
where
\begin{equation}
c_{ij,kl}={}_{AB,\overline A\,\overline B,
\widetilde A\widetilde B}
\braket{ij,ij,zz|C|kl,kl,zz}_{AB,\overline A\,\overline B,
\widetilde A\widetilde B}.
\end{equation}
So, $\mathcal G^\dag(P_0)=0$ and, due to the rule $0\log0=0$, the expressions 
$\mathcal G^\dag\big(\log\mathcal G(\rho)\big)$ and 
$\mathcal G^\dag\big(\log\mathcal Z\big(\mathcal G(\rho)\big)\big)$ are well-defined. 

Also, the elements $\rho_{i,{\rm vac},k,{\rm vac}}$ do not contribute either to the objective function (\ref{EqTarget}) or to the right-hand of Eq.~(\ref{EqGrad}). Hence, even if all these elements are zero, the gradient is still well-defined by continuity (despite that $\rho$ belongs to the boundary of $\mathbf S$). Formally, one can impose some regularization onto the channel $\mathcal G$ (see Eqs.~(12)--(15) in Ref.~\cite{Lutk-numeric}) and then pass to the limit of infinitesimal regularization parameter.

We see that the support of $\mathcal G(\rho_{AB})$ is isomorphic to $\mathbb C^2\otimes\mathbb C^2$.  We can consider $\mathcal G(\rho_{AB})$ to be defined only on the two registers $AB$: the registers $\overline A\,\overline B$ are just copies of $AB$, and the registers $\widetilde A\widetilde B$ contain the fixed value $z$.

For simplicity and graduality, the further proof of Theorem~\ref{Th} will consist of two parts. In the first part we restrict  Bob's Hilbert space to a single-photon subspace spanned by the states $\ket0$ and $\ket1$. Also we put $t=1$ in this case. In the second part we will show that the use of the vacuum component does not give an advantage to Eve. 

\subsection{The case of two-dimensional Bob's space}
 $\mathcal H_B=\mathbb C^2$, and $t=1$.
The matrices $\Gamma_i$ are:
\begin{eqnarray}
\Gamma_1=\eta
&&
\begin{pmatrix}
1&0&0&0\\
0&1&0&0\\
0&0&1&0\\
0&0&0&1
\end{pmatrix},\quad
\Gamma_2=\frac\eta2
\begin{pmatrix}
1&0&0&-1\\
0&1&-1&0\\
0&-1&1&0\\
-1&0&0&1
\end{pmatrix},\nonumber
\\
\Gamma_3=
&&\begin{pmatrix}
1&0&0&0\\
0&\eta &0&0\\
0&0&1&0\\
0&0&0&\eta\label{EqGammas}
\end{pmatrix}
\end{eqnarray}
(the order of rows and columns are as follows: $AB=00,01,10,11$). Note that 
\begin{equation}\label{EqGammaRho}
\Tr\Gamma_i\rho=\sum_{j,k=1}^4\Gamma^{kj}_i\rho_{jk}=
\sum_{j,k=1}^4\big(\Gamma^{jk}_i\big)^*\rho_{jk},
\end{equation}
where $\Gamma^{jk}_i$ and $\rho_{jk}$ are the elements of the corresponding matrices. So, each $\Gamma_i$ fixes a weighted sum of the elements of $\rho$.

Let us prove the necessity of condition (\ref{EqFeas}) for the existence of feasible solutions of the optimization problem. Positive semidefiniteness of $\rho$ implies
\begin{subequations}\label{EqRhoPos}
\begin{eqnarray}
|\rho_{00,11}|&\leq&\sqrt{\rho_{00,00}\rho_{11,11}},\\
|\rho_{01,10}|&\leq&\sqrt{\rho_{01,01}\rho_{10,10}}. 
\end{eqnarray}
\end{subequations}
Denote 
\begin{subequations}\label{EqRhoDiag}
\begin{eqnarray}
\rho_{00,00}&=&(1-Q_z)(1+\delta_0)/2,\\ 
\rho_{11,11}&=&(1-Q_z)(1-\delta_0)/2,\\
\rho_{01,01}&=&Q_z(1-\delta_1)/2,\\ 
\rho_{10,10}&=&Q_z(1+\delta_1)/2.
\end{eqnarray}
\end{subequations}
Then, from  Eqs.~(\ref{EqRestr}), (\ref{EqGammas}), (\ref{EqRhoPos}), and (\ref{EqRhoDiag}) (recall that now $t=1$),
\begin{eqnarray}
2Q_x&=&1-2\Re\rho_{00,11}-2\Re\rho_{01,10}\nonumber\\
&\geq&
1-(1-Q_z)\sqrt{1-\delta_0^2}-Q_z\sqrt{1-\delta_1^2}
\label{EqQx}
\end{eqnarray}
and
\begin{equation}\label{EqPass}
p_{\rm pass}=\frac{1+\eta}2+
\frac{1-\eta}2[(1-Q_z)\delta_0+Q_z\delta_1].
\end{equation}
The right-hand side of Eq.~(\ref{EqQx}) with the restriction (\ref{EqPass}) 
for a fixed $p_{\rm pass}$ takes its minimum for $\delta_0=\delta_1=\delta$. The minimum is equal to the right-hand side of inequality (\ref{EqFeas}), and equality (\ref{EqDelta}) takes place. Hence, inequality (\ref{EqFeas}) is a necessary condition for the existence of a positive semi-definite operator $\rho_{AB}$ satisfying Eq.~(\ref{EqRestr}). 

Consider the following operator:
\begin{eqnarray}
\overline\rho_{AB}=
\frac{1-Q_z}2
&\begin{pmatrix}
1+\delta&0&0&1-2Q_x\\
0&0&0&0\\
0&0&0&0\\
1-2Q_x&0&0&1-\delta
\end{pmatrix}&\nonumber
\\
+
\frac{Q_z}2
&\begin{pmatrix}
0&0&0&0\\
0&1-\delta &1-2Q_x&0\\
0&1-2Q_x&1+\delta&0\\
0&0&0&0
\end{pmatrix}&.\label{EqRhoOpt}
\end{eqnarray}
If inequality (\ref{EqFeas}) is satisfied, then this operator is positive semi-definite and satisfies Eq.~(\ref{EqRestr}) (for $t=1$). This proves the sufficiency of condition (\ref{EqFeas}) for the existence of a solution. As we will see, (\ref{EqRhoOpt}) is an optimal solution. 

Recall that we consider $\mathcal G(\overline\rho_{AB})$ to be defined only on the registers $AB$. Then,
\begin{equation}
\begin{split}
\mathcal G(\overline\rho_{AB})=
\frac{1-Q_z}2
&\begin{pmatrix}
1+\delta&0&0&(1-2Q_x)\sqrt\eta\\
0&0&0&0\\
0&0&0&0\\
(1-2Q_x)\sqrt\eta&0&0&(1-\delta)\eta
\end{pmatrix}
\\
+
\frac{Q_z}2
&\begin{pmatrix}
0&0&0&0\\
0&(1-\delta)\eta&(1-2Q_x)\sqrt\eta&0\\
0&(1-2Q_x)\sqrt\eta&1+\delta&0\\
0&0&0&0
\end{pmatrix},
\end{split}
\end{equation}
and $\mathcal Z\big(\mathcal G(\overline\rho_{AB})\big)$ is  the diagonal part of $\mathcal G(\overline\rho_{AB})$.
The eigenvalues of $\mathcal G(\overline\rho_{AB})$ are:
\begin{equation}
\lambda_{1,2}=(1-Q_z)\lambda_\pm,\quad
\lambda_{3,4}=Q_z\lambda_\pm,
\end{equation}
where $\lambda_-=p_{\rm pass}\lambda(Q_x,\eta,t,p_{\rm pass})$ (see Eq.~(\ref{EqLambda})) and $\lambda_+=p_{\rm pass}-\lambda_-$. Then the straightforward calculation of Eq.~(\ref{EqD}) yields Eq.~(\ref{EqMain}).

\textbf{2. The proof of optimality of} $\overline\rho_{AB}$. We have obtained the desired formula (\ref{EqMain}). It remains to show that $\overline\rho_{AB}$ is optimal. We will show that the gradient $\nabla f(\overline\rho_{AB})$ is orthogonal to  all allowable directions of movement away from $\overline\rho_{AB}$. This will mean that $\overline\rho_{AB}$ provides an optimal value to (\ref{EqRateMaxK}).

Let inequality (\ref{EqFeas}) be satisfied as a strict inequality. In this case, operator (\ref{EqRhoOpt}) is nondegenerate and the gradient, as we concluded above, is well defined. The case when inequality (\ref{EqFeas}) is satisfied as an equality can be obtained as a limiting case. The continuity of the objective function in $\rho_{AB}$ is proved in Ref.~\cite{Lutk-numeric}.

Consider the eigenvectors $(\cos\theta,\sin\theta)$ and $(-\sin\theta,\cos\theta)$, $\theta=\frac12\arctan\frac{2\sqrt\eta(1-2Q_x)}{1-\eta+\delta(1+\eta)}$, of the matrix
\begin{equation}
\begin{pmatrix}
1+\delta&(1-2Q_x)\sqrt\eta\\
(1-2Q_x)\sqrt\eta&(1-\delta)\eta
\end{pmatrix}
\end{equation}
corresponding to the eigenvalues $2\lambda_\pm$. Then the direct calculation according to Eqs.~(\ref{EqGrad}) and (\ref{EqGdag}) gives
\begin{eqnarray}
&&\nabla f(\overline\rho_{AB})
=
\begin{pmatrix}
d_0&0&0&0\nonumber\\
0&\eta d_1&0&0\\
0&0&d_0&0\\
0&0&0&\eta d_1
\end{pmatrix}
\\&&+\sqrt\eta \sin\theta\cos\theta\log\frac{\lambda_+}{\lambda_-}
\begin{pmatrix}
0&0&0&1\\
0&0&1&0\\
0&1&0&0\\
1&0&0&0
\end{pmatrix},\label{EqNabla}
\end{eqnarray}
where
\begin{subequations}
\begin{eqnarray}
d_0&=&\cos^2\theta\log\lambda_++\sin^2\theta\log\lambda_-,\\
d_1&=&\sin^2\theta\log\lambda_++\cos^2\theta\log\lambda_-
-\log\eta.\quad\:
\end{eqnarray}
\end{subequations}

We see that the gradient is orthogonal to all directions except the changes in the diagonal of $\rho_{AB}$ and directions that change the sum of the  secondary diagonal. However, the sum of the secondary diagonal is fixed by the restrictions $\Gamma_1$ and $\Gamma_2$. According to the restrictions $\Gamma_1$ and $\Gamma_3$, the allowable directions of changes in the diagonal $(\Delta\rho_{00},\Delta\rho_{01},\Delta\rho_{10},\Delta\rho_{11})$ satisfy the relations
$\Delta\rho_{00}=-\Delta\rho_{10}$ and
$\Delta\rho_{01}=-\Delta\rho_{11}$.
Hence,
\begin{multline}
\Tr\nabla f(\overline\rho_{AB})\,{\rm diag}(\Delta\rho_{00},\ldots,\Delta\rho_{11})
\\=
d_0(\Delta\rho_{00}+\Delta\rho_{10})+
\eta d_1(\Delta\rho_{01}+\Delta\rho_{11})
=0,
\end{multline}
and the gradient is thus orthogonal to all allowable directions. This means that $\overline\rho_{AB}$ provides a minimum to the secret key rate (\ref{EqRateMaxK}) for the case of two-dimensional Bob's space.

\textbf{3. The case of three-dimensional Bob's space.} Now we return to the case of three-dimensional Bob's space and arbitrary $t\leq1$. Since $t$ is a common factor in all restrictions (\ref{EqRestr}), if we multiply the right-hand side of Eq.~(\ref{EqRhoOpt}) by $t$, then the restrictions will be satisfied for this value of $t$. We can also see the vacuum component of $\rho_{AB}$ does not contribute directly (i.e., besides the factor $t$) either to the secret key rate nor  to the gradient [see Eqs.~(\ref{EqGrad}) and (\ref{EqGdag})] or to the restrictions. So, the state $t\overline\rho_{AB}\oplus0+(1-t)I_2/2\otimes\ket{\rm vac}\bra{\rm vac}$, where $\overline\rho_{AB}$ is a matrix defined by Eq.~(\ref{EqRhoOpt}) on the four-dimensional subspace and $\oplus0$ denotes its embedding into the six-dimensional space $\mathcal H_A\otimes\mathcal H_B=\mathbb C^2\otimes\mathbb C^3$, is an optimal state and the secret key is given by Eq.~(\ref{EqMain}). In other words, the use of the vacuum component and transmission loss do not give an advantage to Eve: Her knowledge per sifted key bit remains the same.

\begin{remark}\label{RemTight}
In Remark~\ref{RemRestr}, we promised to show that the inclusion of additional restrictions does not change  the solution of the optimization problem provided that the natural noise in the channel is described by the depolarizing channel (\ref{EqDep}) and (\ref{EqDepRho}). Indeed, let us consider the maximal set of restrictions $\{\Gamma_{jk}\}$ (see the beginning of  Remark~\ref{RemRestr}) and the corresponding values $\gamma_{jk}=\Tr\Gamma_{jk}\rho_{AB}^0$, where $\rho_{AB}^0$ is given by Eq.~(\ref{EqDepRho}). Then it is straightforward to show that $\overline\rho_{AB}$ with $\delta=0$ satisfies all the restrictions, i.e., $\Tr\Gamma_{jk}\overline\rho_{AB}=\gamma_{jk}$. In other words, $\overline\rho_{AB}$  is a feasible solution. It is also an optimal one: If the gradient (\ref{EqNabla}) is orthogonal to all directions of movement allowed by a certain set of the restrictions, then it is also orthogonal to all directions of movement allowed by a larger set of restrictions.  

In this reasoning, we considered the case $\delta=0$ because this is true whenever the natural noise is described by the depolarizing channel. As we noted after Corollary~\ref{Cor}, a non-zero $\delta$ may be caused by statistical fluctuations. So, analysis of a non-zero $\delta$ is necessary for the development of a finite-key analysis, but it can be set to zero if we are interested in the asymptotic secret key rate.

Then we can consider the following problem: What is a minimal set of restrictions that do not decrease the optimal value of the objective function? Our set $\{\Gamma_1, \Gamma_2, \Gamma_3\}$ is such a choice. The matrices of these operators have simple forms and are directly related to the expression of the gradient $\nabla f(\overline\rho_{AB})$.
\end{remark}

\setcounter{equation}{0}
\setcounter{section}{0}
\renewcommand{\theequation}{B\arabic{equation}}

\smallskip

\section*{Appendix B. Information leakage in the error correction  for the case of detection-efficiency mismatch}\label{SecLeak}

Our aim is to calculate the quantity
$H(\overline A|\overline B\widetilde A\widetilde B)_{\rho^{(4)}}$ in Eq.~(\ref{EqDW}).
Since the registers $\overline A\,\overline B$ are the copies of $AB$ and the registers $\widetilde A\widetilde B$ store the fixed value $z$ in $\rho^{(4)}$ (see the beginning of Appendix~A for details), we can consider $\rho^{(4)}$ to be defined only on the two registers $A$ and $B$: $\rho^{(4)}_{AB}$. Further, $H(A|B)_{\rho^{(4)}}$ is a classical conditional entropy with the two binary random variables $A$ and $B$. It is well known to be upper bounded by $h(Q_z)$, where $Q_z$ is the probability of error ($A\neq B$).

However, in our case, for $\overline\rho_{AB}$ given by Eq.~(\ref{EqRhoOpt}), which corresponds to an optimal Eve's attack, $H(A|B)$ is exactly $h(Q_z)$ whenever $\delta=0$. Indeed the diagonal part of $\overline\rho_{AB}$ takes the form $t(1-Q_z,Q_z,Q_z,1-Q_z)/2$ in this case (the order of the values of the registers is the same as in Appendix~A: $AB=00,01,10,11$). Further, $\rho^{(4)}_{AB}=p_{\rm pass}\mathcal G(\rho_{AB})$. The diagonal part of  $\rho^{(4)}_{AB}$ is then $(1-Q_z,Q_z\eta,Q_z,(1-Q_z)\eta)/(1+\eta)$. This is the joint distribution of $A$. It is straightforward to show that $H(A|B)=h(Q_z)$. So, the state $\overline\rho_{AB}$ simultaneously minimizes the first term in the right-hand side of Eq.~(\ref{EqDW}) and maximizes the second term there. Thus, it minimizes the whole expression (\ref{EqDW}). Hence, we can substitute the second term by its maximal value $h(Q_z)$ without loss of tightness of the bound.

If $\delta\neq0$, then, strictly speaking, $H(A|B)$ is smaller than $h(Q_z)$. However, in this paper we assume that a non-zero value of $\delta$ is caused by the statistical fluctuations or by  Eve's interference of the same order (see the discussion after Theorem~\ref{Th} and Corollary~\ref{Cor}) and, hence, $\delta$ is infinitesimal in the limit $n\to\infty$. Hence, the difference between $H(A|B)$ and $h(Q_z)$ is also infinitesimal.

\end{document}